# Dutt-Khare-Varshni Potential and Its Quantized-by-Heun-Polynomials SUSY Partners as Nontrivial Examples of Solvable Potentials Explicitly Expressible in Terms of Elementary Functions

G. Natanson

ai-solutions Inc.

2232 Blue Valley Dr.

Silver Spring MD 20904 U.S.A.

greg_natanson@yahoo.com

The paper presents a detailed analysis of the so-called 'Double-Root Tangent-Polynomial' (DRtTP) reduction of the regular Gauss-reference (*r*-GRef) potential exactly solvable by hypergeometric functions. It is shown that the Dutt-Khare-Varshni (DKV) potential represents a remarkable particular case when the DRtTP *r*-GRef becomes expressible in terms of elementary functions. This is also true for its *basic* single- or double-step SUSY partners which are either exactly or conditionally exactly quantized by Heun polynomials (Hp-EQ or Hp-CEQ, respectively).

## 1. Introduction

Nearly half a century ago the author [1] found the general expression for the rational potential exactly solvable by hypergeometric functions. One of the most important results of the paper was the proof that the change of variable z(x) converting the one-dimensional Schrödinger equation into the hypergeometric equation satisfies the first-order differential equation

$$(z')^2 = {}_1\wp^{-1}[z(x);\, {}_1T_K] \equiv \frac{4z^2(x)[1-z(x)]^2}{{}_1T_K[z(x)]}, \tag{1.1}$$

where the order K of the so-called [2] 'tangent polynomial' (TP) ${}_1T_K[z]$ may not exceed 2. It was shown that all the known potentials solvable in terms of hypergeometric functions of a *real* argument correspond to the four particular cases when the polynomial $T_2[z]$ has no roots other than 0 or 1, namely,

${}_1T_1[z] = c_1 z$ for the hyperbolic Pöschl-Teller (h-PT) potential [3]; (1.2a)

${}_1T_2[z] = a\, z(z-1)$ for the Darboux-Pöschl-Teller (DPT) potential x) [5, 3]; (1.2b)

${}_1T_0[z] = c_1$ for the Rosen-Morse (RM) potential [6]; (1.2c)

${}_1T_2[z] = c_1 z^2$ for the Eckart/Manning-Rosen (E/MR) potential [7, 8]. (1.2d)

As initially suggested by Gendenshtein [9] and directly confirmed by Dutt, Khare, and Sukhatme [11] several years later all four potentials turned out to be 'shape-invariant' xx), namely, they

---

x) As pointed to more recently by Matveev [4] both eigenfunctions and eigenvalues for the appropriate Sturm-Liouville (SL) problem were originally obtained by Darboux [5] long before the birth of the Schrödinger quantum mechanics.

xx) In the original (Russian) versions of Gendenshtein's papers [9, 10] he uses the Russian equivalent of the English term 'form-invariance', retained in the translation of his joint paper with Krive [10]. The broadly accepted term: 'shape-invariance' first appeared in the English translation of [9].

retain their analytical form under Darboux transformations (DTs) with nodeless normalizable factorization fractions (FFs). As pointed to in [2] any rational SUSY partner of the generic Gauss-reference (GRef) has poles at the TP zeros so that the aforementioned common feature of the shape-invariant regular GRef ('*r*-GRef') potentials is the direct consequence of the fact that the TPs used for their construction have no roots other than 0 or 1 (if any).

The author's observation [1] that potentials defined by TPs (1.2a)-(1.2d) can be expressed in terms of elementary functions is often incorrectly interpreted as an intrinsic feature of their shape invariance. One of the purposes of this paper is to address Gomez-Ullate, Grandati, and Milson's [12] recent statement that that all "known potentials for which it is to be possible to determine explicitly all the bound states in terms of elementary transcendental functions and to write explicitly their energies in terms of the quantum number" belong to category of 'shape-invariant' potentials. Indeed this is true that eigenfunctions of the Schrödinger equation with GRef potentials cannot be generally expressed in terms of elementary functions. Below we however analyze a counter-example of the *r*-GRef potential which is expressible in terms of elementary functions despite the fact that it does not preserve its form under the DT with nodeless normalizable FF. Namely, we prove that the Dutt-Khare-Varshni (DKV) potential [12] conventionally treated [14] (in following [13]) as 'conditionally exactly solvable' is nothing but the particular case of the Double-Root TP (DRtTP) potential [2] and therefore it is *exactly* quantized (not conditionally exactly quantized 0by Jacobi polynomials (Jp-EQ). (Here we refer to the rational Liouville potential as 'exactly quantized' if potential parameters can be independently varied without affecting positions of its singular points.)

It is worth mentioning in this context that this work was influenced to great extent by the already cited paper of Roychoudhury et al [14] which demonstrated that the first-order differential equation

$$\hat{\eta}' = \frac{C[1-\hat{\eta}^2(x)]}{\hat{\eta}(x)} < 0 \tag{1.3}$$

for the variable $1 < \hat{\eta}(x) < \infty$ can be integrated via elementary functions of x. It will be shown below that the variable

$$z(x) = \frac{2}{\hat{\eta}(x) + 1} \tag{1.4}$$

satisfies the first-order differential equation

$$(z')^2 = \frac{4z^2(1-z)^2}{(2-z)^2} \tag{1.5}$$

so that the appropriate *r*-GRef potential can be generated by means of the TP with the double root (DRt) of 2. We will refer to this potential as ‘elementary DRtTP’ (*e*-DRtTP) since it can be expressed and therefore quantized in terms of elementary functions.

As a rule [2] the energy spectrum of any DRtTP *r*-GRef potential can be found by solving a cubic equation. However this cubic equation can be analytically decomposed in the limiting case of asymptotically levelled (AL) potential curves [2]. Since this special case was originally discovered in [15] we refer to it as the Williams-Levai (WL) potential. It directly follows from our analysis that eigenfunctions for the elementary WL (e-WL) potential constructed by means of the TP with the DRt $z_T = 2$ are expressible in terms of elementary functions and therefore the same is true for its multi-step rational SUSY partners.

The paper is composed as follows. In Section 2 we review main features of Gauss-seed (GS) solutions for the LW potential. The reason for doing it is that (contrary to the conventional radical formulas for roots of the quadratic equation) even real zeros of the cubic equation require computation of complex roots of complex quantitates which makes it very difficult to analytically determine how many GS solutions the given RCSLE has. It is much easier to take advantage of the general procedure developed by us in [2] for the generic *r*-GRef potential, namely, we start from the AL potential curve and then monotonically increase the reflective barrier at $-\infty$. Similarly to the analysis presented in [2] for the *r*-GRef potential constructed by means of the TP with positive discriminant we prove in Section 3 that the $n^{th}$ eigenfunction $\mathbf{c}_n$ is accompanied by two GS solutions $\dagger_n$ of distinct types. One and only one of these two solutions is necessarily regular at either z=0 or z=1 ($\dagger = \mathbf{a}$ or $\mathbf{b}$ depending on sign of the DRt $z_T$).

In Section 4 we focus more specifically on the explicit *r*-GRef potential constructed by means of the TP with the DRt $z_T = 2$ and in particular directly relate our approach to the more specific expressions derived by Roychoudhury et al [14] for the DKV potential.

It directly follows from the comprehensive theory developed in [2] that the canonical Liouville-Darboux transformation (CLDT) of the DRtTP *r*-GRef potential using one of three basic FFs †0 results in a rational potential exactly quantized by Heun polynomials (Hp-EQ). And in Section 5 we use the related Heun equations to illustrate the general formalism.

As already pointed to in [16] and discussed in more detail in Section 6, the double-step canonical Liouville-Darboux transformation (CLDT) of the *r*-GRef potential using a pair of basic GS solutions †0 and †′0 leads to a rational potential conditionally exactly quantized by Heun polynomials (Hp-CEQ). A trivial combination of GS solutions includes a ground-energy eigenfunction **c**0 and a regular basic solution (which necessarily lies below the ground energy level). In addition we prove in Appendix A that there is a range of the ray identifiers (RIs) where the regular basis GS solution lies below the irregular basic solution **d**0 so that this pair of basic GS solutions can be also used to construct the second Hp-CEQ potential. In Appendix B we analyze some specific features of the Heun equation for the FFs †0;†′0 used to construct both double-step Hp-CEQ potentials.

Finally Section 7 briefly outlines the most important results of the paper as well as its future extensions to multi-step CLDTs with regular Gauss seed solutions †m below ground-energy level coupled with pairs **c**,v;**c**,v+1 of juxtaposed eigenfunctions.

## 2. **Algebraic formulas for energies of AEH solutions for the Williams-Levai potential**

To match Williams and Levai's notation to our general convention for GRef potentials [2] it is convenient to introduce an auxiliary variable $\eta(\breve{x})$ by defining it via first-order differential equation (12) in [15]:

$$\frac{d\eta}{d\breve{x}} = \frac{1-\eta^2}{\eta+\gamma} \qquad (-1 < \eta < +1). \tag{2.1}$$

We then change z for η in all the relevant formulas in [15]. The WL potential can be thus represented as

$$V_{WL}[\eta] = -\frac{A(1-\eta^2)}{4(\eta+\gamma)^2} - \frac{\eta(1-\eta^2)}{(\eta+\gamma)^3} - \frac{3}{4}\frac{(1-\eta^2)^2}{(\eta+\gamma)^4}. \tag{2.2}$$

The DRt TP in the denominator of (1.1) can be conveniently written as

$$T_2[z; z_T] = (z - z_T)^2 / (1 - z_T)^2 \tag{2.3}$$

by requiring that

$$_1c_1 \equiv T_2[1; z_T] = 1. \tag{2.4}$$

The change of variable

$$\eta(x;\eta_T) = 2z(x;z_T) - 1 \qquad (\eta_T \equiv 2z_T - 1) \tag{2.5}$$

then leads to the differential equation

$$\eta' = \frac{(1-\eta^2)(1-\eta_T)}{\eta - \eta_T} \tag{2.6}$$

which turns into (2.1) if we set

$$\breve{x} = \breve{x}_T x, \tag{2.7}$$

$$\breve{x}_T \equiv \eta_T - 1 \equiv 2(z_T - 1), \tag{2.7'}$$

and $\eta_T = -\gamma$. As pointed to in [15] one can always assume that the parameter $\gamma$ is larger than 1 by changing η and $\gamma$ respectively for −η and $-\gamma$ otherwise. It would be however shown in next Section that there are two different branches of the DRtTP potential: $\eta_T < -1$ $(z_T < 0)$ and $\eta_T > +1$ $(z_T > 1)$ for nonzero values of the reflective barrier. Both braches merge into the WL potential in the AL limit.

Representing the derivative of z with respect to $\breve{x}$ as

$$\frac{dz}{d\breve{x}} = \frac{z_T(1-z_T) - z(1-z)}{z - z_T} - \frac{z_T(1-z_T)}{z - z_T} \tag{2.8}$$

and taking advantage of the alternative formula for the Schwartz derivative [17]

$$\{z, x\} = (z''/z')' - \tfrac{1}{2}(z''/z')^2, \tag{2.9$'$}$$

instead of its conventional definition

$$\{z, x\} = z'''/z' - \tfrac{3}{2}(z''/z')^2 \tag{2.9}$$

initially adopted by us [1] from Bose's paper [18], we find

$$\breve{x}_T^{-2}\{z, x\} = \{z, \breve{x}\} = -\frac{1}{2(z - z_T)^2} + \frac{z(1-z)(2z-1)}{(z - z_T)^3} + \frac{3z^2(1-z)^2}{2(z - z_T)^4}. \tag{2.10}$$

By combining the latter expression with the generic formula for the reference polynomial fraction (RefPF)

$$_1I^o[z; \lambda_o, \mu_o] = \frac{1-\lambda_0^2}{4z^2} + \frac{1}{4(1-z)^2} + \frac{\mu_o^2 - \lambda_o^2 + 1}{4z(1-z)} \tag{2.11}$$

we can thus represent the AL reduction of the DRtTP *r*-GRef potential as

$$\breve{x}_T^{-2} V[z \,|\, {}_1\breve{\boldsymbol{G}}^{210}(z_T)] = \frac{(1-\mu_0^2)z(1-z)}{(z - z_T)^2} - \frac{z(1-z)(2z-1)}{2(z - z_T)^3} - \frac{3z^2(1-z)^2}{4(z - z_T)^4}. \tag{2.11*}$$

using the classification scheme [2]:

K= TP order,

$\mathfrak{I}$ = number of outer roots,

$\aleph$ = order of zero root

for a 'regular' polynomial fraction beam ($r$-PFB) ${}_1\mathbf{B}^{K\Im\aleph}$ of our current interest. (In following [2] we write the symbol $\cup$ on **B** to indicate that we deal with the AL-reduction of the generic PFB.) Each PFB ${}_1\breve{G}^{210}(z_T)$ is formed by the Bose [18] invariants I[z;ε;$\mu_o$;$z_T$] dependent on a single ray identifier (RI) $\mu_o$ which is varied, together with the energy ε at the fixed value of the DRt $z_T$.

As a direct result of the arguments presented in [2], we find

$$V[0\,|\,{}_1\breve{G}^{210}(z_T)] = V[1\,|\,{}_1\breve{G}^{210}(z_T)] = 0. \tag{2.12}$$

Substituting (2.5) into (2.2) and comparing the resultant expression with (2.12) shows that

$$V_{WL}[2z-1] = \breve{x}_T^{-2} V[z\,|\,{}_1\breve{G}^{210}(z_T)], \tag{2.13}$$

with

$$\tfrac{1}{4}A = \mu_o^2 - 1 \equiv f_o. \tag{2.14}$$

The common remarkable feature of the AL potential curves is that that energies of AEH solutions can be written in an explicit form. In general (K=$\Im$=2) there are four co-existent AEH solutions **†**m of distinct type (**†** = **a**, **b**, **c**, and **d**)

$$\phi_{\dagger m}[z\,|\,{}_1G^{K\Im 0}_{\downarrow \dagger m}] = z^{{}_1\rho_{0;\dagger m}}(1-z)^{{}_1\rho_{1;\dagger m}}\,\Pi_m[z;\bar{z}_{\dagger m}] \tag{2.15}$$

$$\equiv z^{\frac{1}{2}({}_1\lambda_{0;\dagger 0}+1)}(1-z)^{\frac{1}{2}({}_1\lambda_{1;\dagger 0}+1)}\,\Pi_m[z;\bar{z}_{\dagger m}], \tag{2.15*}$$

where $\Pi_m[z;\bar{z}_{\dagger m}]$ is a polynomial of order m with m distinct zeros $\bar{z}_{\dagger m}$. The important advantage of using the signed exponent differences (signed ExpDiffs)

$${}_1\lambda_{r;\dagger m} = 2\,{}_1\rho_{r;\dagger m} - 1, \tag{2.16}$$

instead of the characteristic exponents (ChExps) ${}_1\rho_{r;\mathbf{t}\mathrm{m}}$, is that their absolute values are related via the quadratic formula

$$ {}_1\lambda^2_{0;\mathbf{t}\mathrm{m}} = \lambda^2_{\mathrm{o}} + {}_1c_0\,{}_1\lambda^2_{1;\mathbf{t}\mathrm{m}} \tag{2.17}$$

as far as the zero-point energy for the *r*-GRef potential is chosen via the requirement $V[1\,|\,{}_1\boldsymbol{G}] = 0$ so that

$$ {}_1\varepsilon_{\mathbf{t}\mathrm{m}} = -\,{}_1\lambda^2_{1;\mathbf{t}\mathrm{m}} \tag{2.18}$$

if ${}_1c_1$ is set to 1. In the limiting case of the AL potential curves ($\lambda_{\mathrm{o}} = 0$) quadratic formula (2.17) turns into the linear relation

$$ {}_1\breve{\lambda}_{0;\breve{\mathbf{t}}_{\pm}\mathrm{m}} = \pm\sqrt{{}_1c_0}\;{}_1\breve{\lambda}_{1;\breve{\mathbf{t}}_{\pm}\mathrm{m}}, \tag{2.19}$$

where

$$ \breve{\mathbf{t}}_{+} = \mathbf{c} \text{ or } \mathbf{d} \tag{2.20$^{+}$}$$

and

$$ \breve{\mathbf{t}}_{-} = \mathbf{a} \text{ or } \mathbf{b} \tag{2.20$^{-}$}$$

We [2] refer to the sequence of AEH solutions of the given type $\mathbf{t}$ as ‘primary’ if it starts from the basic solution $\mathbf{t}0$. It has been proven in [2] that the number of AEH solutions in each primary sequence is equal to or exceeds the number ${}_1n_{\mathrm{o}}$ of bound states if the TP has a positive discriminant. Under this constraint there are exactly four primary sequences: $\mathbf{a}$m, $\mathbf{b}$m, $\mathbf{c}$v (v = 0, 1,…, $v_{max} = {}_1n_{\mathrm{o}} - 1$), and $\mathbf{d}$m. As for supplementary sequences each of symbols $\mathbf{a}$, $\mathbf{b}$, and $\mathbf{d}$ will be accompanied by one or possibly even by two primes.

Only three of the four primary sequences retain in the DRtTP limit. In fact, it directly follows from (5.1.21) and (5.1.21′) in [2] that the leading coefficient of the polynomial $-\breve{G}^{(m)}_2[\lambda\,|\,{}_1\breve{\boldsymbol{G}}^{2\Im 0}]$ vanishes in the limiting case of the TP with zero discriminant ($\Im$=1) so that the appropriate quadratic equation for signed exponent differences turns into the linear formula

$$ {}_1\breve{\lambda}_{1;\breve{\mathbf{t}}_m} = \frac{g_m(\mu_o)}{2(\sqrt{{}_1c_0}-1)(2m+1)}, \tag{2.21} $$

where

$$ \breve{\mathbf{t}}_ = \begin{cases} \mathbf{a} \ \text{ for } \mu_o > 2m+1 \ (m < {}_1n_o) \text{ and } {}_1c_0 > 1, \\ \mathbf{b} \ \text{ for } \mu_o > 2m+1 \ (m < {}_1n_o) \text{ and } {}_1c_0 < 1, \\ \mathbf{a}' \ \text{ for } 2m+1 > \mu_o \ (m \ge {}_1n_o) \text{ and } {}_1c_0 < 1, \\ \mathbf{b}' \ \text{ for } 2m+1 > \mu_o \ (m \ge {}_1n_o) \text{ and } {}_1c_0 > 1, \end{cases} \tag{2.21$'$} $$

and

$$ g_m(\mu_o) \equiv (2m+1)^2 - \mu_o^2. \tag{2.22} $$

Substituting

$$ \frac{z_T}{z_T - 1} = \sqrt{{}_1c_0} > 0 \tag{2.23} $$

into (2.21), coupled with (2.18), thus gives

$$ {}_1\breve{\varepsilon}_{\breve{\mathbf{t}}_m} = -\frac{(z_T-1)^2 g_m^2(\mu_o)}{4(2m+1)^2} \tag{2.24} $$

An analysis of inequality (2.23) also shows that

$$ z_T < 0 \text{ if } {}_1c_0 < 1 \tag{2.25a} $$

and

$$ z_T > 1 \text{ if } {}_1c_0 > 1. \tag{2.25b} $$

The bound energies ${}_1\breve{\varepsilon}_{\mathbf{c}0} = -{}_1\breve{\lambda}^2_{1;\mathbf{c}0}$ are given by positive roots of the quadratic equation

$$ 4\sqrt{{}_1c_0}\,{}_1\breve{\lambda}^2_{1;\breve{\mathbf{t}}_+m} + 2(\sqrt{{}_1c_0}+1)(2m+1)\,{}_1\breve{\lambda}_{1;\breve{\mathbf{t}}_+m} + g_m(\mu_o) = 0 \tag{2.26} $$

with a nonnegative discriminant

$$ {}_1\breve{\Delta}_m(\mu_o;{}_1c_0) = 4[(2m+1)^2(\sqrt{{}_1c_0}-1)^2 + 4\sqrt{{}_1c_0}\,\mu_o^2\,] \ge 0. \tag{2.27} $$

Similarly to the generic case of the *r*-GRef potential with a nonzero reflective barrier the root positive for m smaller than the number of bound energy states

$$ {}_1n_o \equiv [\tfrac{1}{2}(\mu_o - 1)] \tag{2.28}$$

changes its sign for $m \geq n_o$. However, the m[th] eigenfunction now turns into the AEH solution $\mathbf{d}'m$ (not $\mathbf{a}'m$) for $2m+1>\mu_o$, i.e.,

$$ {}_1\breve{\lambda}_{1;\breve{\mathbf{t}}_+ m} = \frac{-2(\sqrt{{}_1c_0}+1)(2m+1)+\sqrt{{}_1\breve{\Delta}_m(\mu_o;{}_1c_0)}}{8\sqrt{{}_1c_0}} \tag{2.29}$$

for

$$ \breve{\mathbf{t}}_+ = \begin{cases} \mathbf{c} & \text{if } \mu_o > 2m+1, \\ \mathbf{d}' & \text{if } \mu_o < 2m+1. \end{cases} \tag{2.29$'$}$$

Here we mark $\mathbf{d}$ by prime to distinguish this AEH solution from the one in the primary sequence. In the latter case the appropriate negative root of quadratic equation (2.20) is given by the same formula

$$ {}_1\breve{\lambda}_{1;\mathbf{d}m} = \frac{-2(\sqrt{{}_1c_0}+1)(2m+1)-\sqrt{{}_1\breve{\Delta}_m(\mu_o;{}_1c_0)}}{8\sqrt{{}_1c_0}} \tag{2.29*}$$

regardless of the value of $\mu_o$. It should be stressed in this connection that the arguments presented in [2] were explicitly based on the assumption that the RI $\lambda_o$ has a positive value. We postpone further discussion of this nontrivial behavior of eigenfunctions in the limiting case $\lambda_o = 0$ util next Section which presents a detailed analysis of AEH solutions for the generic DRtTP potential.

Representing quadratic equation (2.26) as

$$ (\gamma-1)\,{}_1\breve{\lambda}^2_{1;\breve{\mathbf{t}}_+ m} + \gamma(2m+1)\,{}_1\breve{\lambda}_{1;\breve{\mathbf{t}}_+ m} + (\gamma+1)g_m(\mu_o) = 0 \tag{2.30}$$

we conclude that two roots given by (20) in [15] are nothing but signed Exp-Diffs (2.29) and (2.29*). In particular, if the upper sign is selected we come to Williams and Levai formula (18) for bound energy levels:

$$a^2 = -\breve{x}_T^{-2}\, {}_1\breve{\varepsilon}_{\mathbf{c}n}. \tag{2.31}$$

Now we come to the most challenging part of this section – explicit constraints selecting all the regular AEH solutions below the ground energy level ${}_1\breve{\varepsilon}_{\mathbf{c}0}$. Let us start from AEH solutions of types $\breve{\mathbf{t}}_-$. The exponent differences $|{}_1\breve{\lambda}_{1;\breve{\mathbf{t}}_-m}|$ below the upper bound ${}_1\lambda_{1;\mathbf{c}0}$ are determined by the following constraints on m:

$$^{+}\breve{P}_2^{\mathbf{c}0}(m;{}_1c_0) \equiv (2m+1)^2 + 2\,|\sqrt{{}_1c_0}-1|\,{}_1\breve{\lambda}_{1;\mathbf{c}0}(2m+1)] - \mu_o^2 < 0 \quad (m < {}_1n_o) \tag{2.32}$$

and

$$^{-}\breve{P}_2^{\mathbf{c}0}(m;{}_1c_0) \equiv (2m+1)^2 - 2\,|\sqrt{{}_1c_0}-1|\,{}_1\breve{\lambda}_{1;\mathbf{c}0}(2m+1)] - \mu_o^2 > 0 \quad (m \geq {}_1n_o) \tag{2.32′}$$

for the primary and supplementary sequences accordingly. As expected the first set of the regular solutions includes the basic solution $\breve{\mathbf{t}}_-0$. Since each polynomial in the left-hand side of (2.32) and (2.32′) has a positive leading coefficient and a negative free term its roots must have an opposite sign so that the upper bound for nodeless regular AEH solutions from the primary sequence and the lower bound for nodeless AEH solutions from the supplementary sequence must be smaller and respectively larger than the positive zeros ${}^{+}m_{\mathbf{c}0}(\mu_o;{}_1c_0)$ and ${}^{-}m_{\mathbf{c}0}(\mu_o;{}_1c_0)$ of polynomials (2.32) and (2.32′) expressed in terms of m. An analysis of the inequality

$$2\,{}^{+}m_{\mathbf{t}_+0}(\mu_o;{}_1c_0)+1 \equiv \frac{\mu_o^2}{|\sqrt{{}_1c_0}-1|\,{}_1\lambda_{1;\mathbf{t}_+0}+\sqrt{(\sqrt{{}_1c_0}-1)^2\,{}_1\lambda^2_{1;\mathbf{t}_+0}+\mu_o^2}} < \mu_o \tag{2.33}$$

shows that that ${}^{+}m_{\mathbf{c}0}(\mu_o;{}_1c_0) < {}_1n_o$ so that the primary sequence $\breve{\mathbf{t}}_-m$ (m = 0, …, ${}_1n_o-1$) contains exactly $[{}^{+}m_{\mathbf{c}0}(\mu_o;{}_1c_0)]$ nodeless AEH solutions (m = 0, 1,…, $[{}^{+}m_{\mathbf{c}0}(\mu_o;{}_1c_0)]-1$).

Similarly the inequality

$$2\,\bar{m}_{\mathbf{t}_+0}(\mu_o;{}_1c_0)+1 \equiv |\sqrt{{}_1c_0}-1|\,{}_1\lambda_{1;\mathbf{t}_+0}+\sqrt{(\sqrt{{}_1c_0}-1)^2\,{}_1\lambda^2_{1;\mathbf{t}_+0}+\mu_o^2} > \mu_o \qquad (2.33')$$

shows that that the infinite subsequence of nodeless regular AEH solutions $\breve{\mathbf{t}}_m$ starts from

$m = [\bar{m}_{\mathbf{c}0}(\mu_o;{}_1c_0)]+1$.

As a direct consequence of the general proof presented in [2] the AEH solutions $\mathbf{a}m$ and $\mathbf{d}m$ for $\mu_o > 2m+1$ co-exist at the same energy $\varepsilon = -(2m+1)^2$ at the crossing point

$$\mu_o = \mu_{\times;m} \equiv \sqrt{2\sqrt{{}_1c_0}-1}\,(2m+1) \qquad (2.34)$$

between the $\mathbf{a}/\mathbf{d}$-DRt hyperbola

$$h_m(\lambda_o,\mu_o;-2\sqrt{{}_1c_0}) \equiv \mu_o^2-\lambda_o^2+(1-2\sqrt{{}_1c_0})(2m+1)^2 = 0 \qquad (2.35)$$

and the straight line $\lambda_o = 0$. This crossing point exists only if ${}_1c_0 > 1$ simply because there is no AEH solution of type $\mathbf{a}$ in this case for $\mu_o > 2m+1$.

## 3. The generic DRtTP *r*-GRef potential as a rational extension of the WL potential toward positive reflective barriers

It has been proven in [2] that the energies ${}_1\varepsilon_{\mathbf{t}m} = -{}_1\lambda^2_{1;\mathbf{t}m}$ of AEH solutions for the DRtTP potential

$$x_T^{-2}V[z\,|\,{}_1\mathcal{G}^{210}(z_T)] = \frac{[\lambda_o^2(1-z)-f_o\,z](1-z)}{(z-z_T)^2}-\frac{z(1-z)(2z-1)}{2(z-z_T)^3}-\frac{3z^2(1-z)^2}{4(z-z_T)^4} \qquad (3.1)$$

are unambiguously determined by real zeros of the cubic polynomial $G_3^{(m)}[\lambda\,|\,{}_1\mathcal{G}^{210}]$ with the coefficients

$$ {}_1G^{(m)}_{3;3}({}_1c_0) \equiv {}_1G_m(c_0) \equiv 8\sqrt{c_0}(1-\sqrt{c_0})(2m+1), \tag{3.2a} $$

$$ {}_1G^{(m)}_{3;2}(\lambda_o,\mu_o\,;{}_1c_0) = -4\{\sqrt{c_0}[\mu_o^2-\lambda_o^2-(2m+1)^2]+\lambda_o^2+({}_1c_0-1)(2m+1)^2\}, \tag{3.2b} $$

$$ {}_1G^{(m)}_{4;1}(\lambda_o,\mu_o) = {}_1G^{(m)}_{3;1}(\lambda_o,\mu_o) = -4(2m+1)[\mu_o^2+\lambda_o^2-(2m+1)^2], \tag{3.2c} $$

$$ {}_1G^{(m)}_{3;0}(\lambda_o,\mu_o) = [\mu_o^2-(\lambda_o-2m-1)^2]\times[\mu_o^2-(\lambda_o+2m+1)^2] \tag{3.2d} $$

$$ = [(2m+1)^2-(\lambda_o-\mu_o)^2]\times[(2m+1)^2-(\mu_o+\lambda_o)^2] \tag{3.2d*} $$

In previous Section we simply took advantage of the fact that this polynomial can be analytically decomposed into the product of the linear and quadratic polynomials in the AL limit ($\lambda_o$= 0).

Since the variable z(x) was chosen in such a way that potential (3.1) vanish for z=1 one has to consider two branches $z_T > 1$ (${}_1c_0 > 1$) and $z_T < 0$ (${}_1c_0 < 1$) separately. In the limit $|z_T| \to \infty$ (${}_1c_0 \to 1$) both branches collapse into the Rosen-Morse potential.

Though one can formally write down radical expressions for sought-for real roots of the given cubic polynomial the analysis of the appropriate formulas seems to be too cumbersome to further proceed in this direction. A much easier procedure outlined for the generic *r*-GRef potential in [2] is to start from the AL potential curves (the WL potential in our case) and then to monotonically increase the reflective barrier. It was proven that the m$^{th}$ eigenfunction is always accompanied by a pair of AEH solutions $\breve{\dagger}_{+}m$ and $\breve{\dagger}_{-}m$ of two distinct types so that two real roots of the cubic polynomials cannot merge into a double root for $\lambda_o = 0$ as far as the potential has at least m+1 bound energy levels.

Since the free term of quartic polynomial (5.1.20a) in [2] is independent of TP coefficients AEH solutions $\dagger m$ may change their type only along the straight lines

$$ \mu_o = \lambda_o + 1, \tag{3.3a} $$

$$ \mu_o + \lambda_o = 1, \tag{3.3b} $$

$$\mu_o + 1 = \lambda_o, \tag{3.3c}$$

as in the generic case analyzed in [2]. These 'zero-factorization energy' (ZFE) separatrices (as we refer to them in [2]) split the $\lambda_o\ \mu_o$ plane into four major areas depicted in Figs. 1 and 2 for m=2. Namely, the central region carved by three ZFE separatrices $A_m|D_m$, $B_m|D_m$, or $C_m|D_m$ is marked as Area $D_m$. Each ZFE separatrix (3.3a), (3.3b), or (3.3c) serves as the border line between this central region and Area $A_m$, $B_m$, or $C_m$, respectively, i.e.,

Area $A_m$: $\mu_o > \lambda_o + 2m + 1;$ (3.4a)

Area $B_m$: $\mu_o + \lambda_o < 2m + 1;$ (3.4b)

Area $C_m$: $\mu_o < \lambda_o - 2m - 1;$ (3.4c)

Area $D_m$: otherwise. (3.4d)

The arrows in both figures indicate that the presented classification of the triplets has been proven only for the neighborhood of the given border line. The solid lines depict the major ($A_0|D_0$) $\mathbf{c}/\mathbf{a}'$ separatrices which split the $\lambda_o\ \mu_o$ plane into two sub-domains with and without the discrete energy spectrum accordingly.

Note that free term (3.2d*) is positive for $2m + 1 < \mu_o - \lambda_o$ so that the polynomial has an even (odd) number of negative roots for $_1c_0 > 1$ ($_1c_0 < 1$) in Area $A_m$ in agreement with the results of previous Section for the zero reflective barrier ($\lambda_o = 0$).

By making use of the fractional relation [2]

$$_1\lambda_{0;\dagger m} = \frac{\mu_o^2 - \lambda_o^2 + (1 - 2\sqrt{_1c_0})\,_1\lambda_{1;\dagger m}^2 - (_1\lambda_{1;\dagger m} + 2m + 1)^2}{2(_1\lambda_{1;\dagger m} + 2m + 1)} \tag{3.5}$$

between signed Exp-Diffs $_1\lambda_{1;\dagger m}$ and $_1\lambda_{0;\dagger m}$ one finds

$$_1\lambda_{0;\dagger m}\Big|_{_1\varepsilon_{\dagger m}=0} = \begin{cases} \lambda_o \text{ along } A_m \mid D_m, \\ -\lambda_o \text{ along } B_m \mid D_m, \\ -\lambda_o \text{ along } C_m \mid D_m \end{cases} \tag{3.6}$$

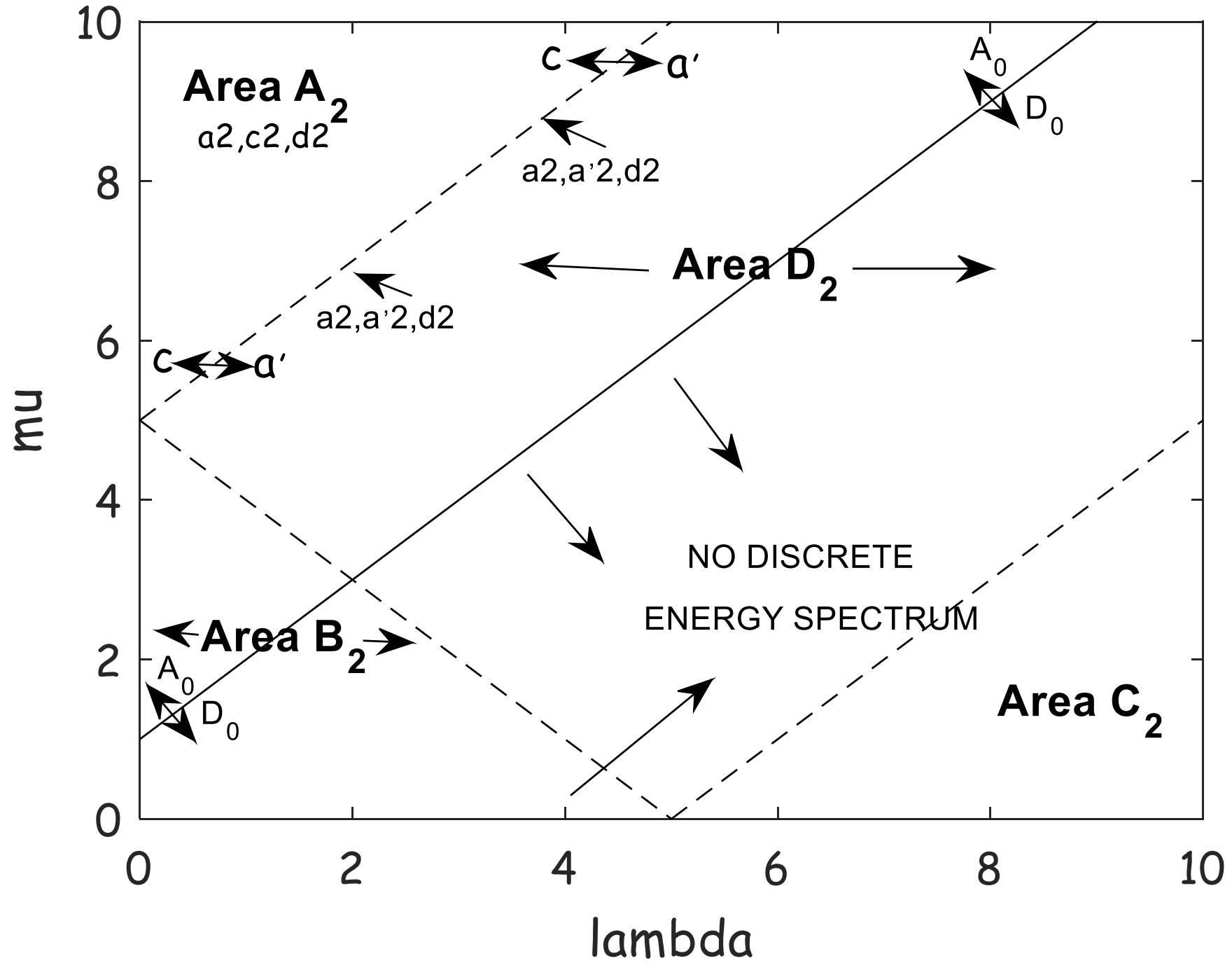

**Figure 1. Triplets of AEH solutions †2 of a distinct type in Areas $A_2$ of the $\lambda_o\ \mu_o$ plane for the TP with a positive outer zero ($_1c_0 > 1$).**

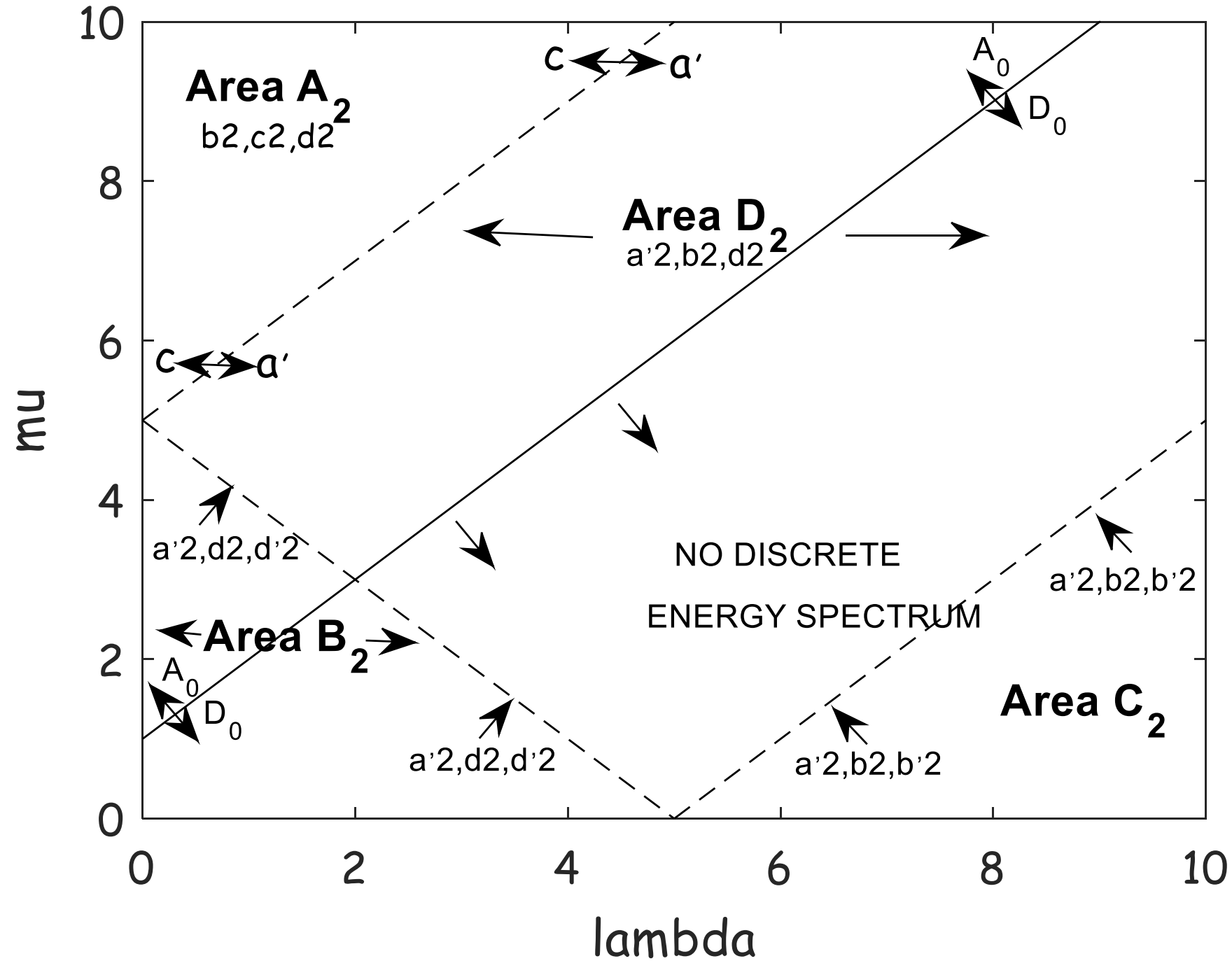

**Figure 2. Triplets of AEH solutions †2 of a distinct type in Areas $A_2$ and $D_2$ of the $\lambda_o\ \mu_o$ plane for the TP with a negative outer zero ($_1c_0 < 1$).**

so that zero-energy AEH solutions may come only in pairs of types **c** and **d** along the ZFE separatrix $A_m|D_m$ or of types **b** and **d** along two other border lines. In particular this implies (in agreement with the general theory [2]) that the eigenfunction **c**m turns into the AEH solution **a′**m on other side of the border line $A_m|D_m$ (the so-called 'm$^{th}$ **c**/**a′** separatrix' in terms of [2]). As for two other ZFE separatrices $B_m|D_m$, or $C_m|D_m$ the AEH solutions may only change their type either from **b** to **d** or from **d** to **b** depending on the direction of the border crossing.

It has been also proven in [2] that Area $A_m$ represents the range of RIs $\mu_o$ and $\lambda_o$ where the *r*-GRef potential has at least m+1 bound energy levels. For TP with positive discriminant $(\Delta_T > 0)$ each eigenfunction **c**m is accompanied by three AEH solutions †m of three distinct types **a**, **b**, and **d**. Only two of these AEH solutions survive in the DRtTP limiting case. As shown in previous Section the eigenfunction **c**m in the limiting case of the WL potential ($\lambda_o = 0$) is accompanied by a pair of AEH solutions **a**m and **d**m if ${}_1c_0 >1$ or **b**m and **d**m if ${}_1c_0 >1$.

To independently confirm the latter assertion let us analyze DRtTP reduction

$$\begin{aligned}\tilde{G}_3^{(m)}[\lambda \mid {}_1\boldsymbol{G}^{210}] &\equiv {}_1\tilde{G}_3^{(m)}[\lambda;\lambda_o,\mu_o;{}_1c_0] \\ &= {}_1c_0[\mu_o^2-(\lambda+2m+1)^2+(1-2/\sqrt{{}_1c_0})(\lambda^2-\lambda_o^2)]^2 \\ &\quad - 4(\lambda^2-\lambda_o^2)(\lambda+2m+1)^2 \end{aligned} \tag{3.7}$$

of generic quartic polynomial

$$\begin{aligned}\tilde{G}_4^{(m)}[\lambda \mid {}_1\boldsymbol{G}] \equiv {}_1c_0[\mu_o^2-(\lambda+2m+1)^2+(\lambda^2-\lambda_o^2)({}_1a_2-1)/{}_1c_0]^2 \\ - 4(\lambda^2-\lambda_o^2)(\lambda+2m+1)^2 \end{aligned} \tag{3.8}$$

determining signed ExpDiffs ${}_1\lambda_{0;\dagger m}$ via the quartic equation

$$\tilde{G}_4^{(m)}[{}_1\lambda_{0;\dagger m} \mid {}_1\boldsymbol{G}] = 0. \tag{3.8*}$$

One can easily verify the latter equation by starting from the conditions

$$\mu_{\dagger m} \equiv {}_1\lambda_{0;\dagger m} + {}_1\lambda_{1;\dagger m} + 2m+1 \tag{3.9*}$$

and

$$\mu^2_{\dagger m} = \mu^2_o + {}_1a_2\, {}_1\lambda^2_{1;\dagger m} \tag{3.9}$$

(used in [2] as the starting point to derive the quartic equation for ${}_1\lambda_{1;\dagger m}$) and then excluding ${}_1\lambda_{1;\dagger m}$ (instead of ${}_1\lambda_{0;\dagger m}$) via (2.17). Note that polynomial (3. 8) has a positive free term (except the point $\lambda_o = 0$, $\mu_o = 2m+1$ where it vanishes) and therefore this must be also true for its cubic reduction (3.7) with the leading coefficient

$${}_1\tilde{G}^{(m)}_{3;3}({}_1c_0) \equiv {}_1\tilde{G}_m({}_1c_0) = 8(\sqrt{{}_1c_0} - 1)(2m+1). \tag{3.10}$$

and the free term

$${}_1\tilde{G}^{(m)}_{3;0}(\lambda_o, \mu_o; {}_1c_0) = \{(2\sqrt{{}_1c_0} - 1)\lambda^2_o + {}_1c_0[\mu^2_o - (2m+1)^2]\}^2 + 4\lambda^2_0(2m+1)^2 \tag{3.10$_0$}$$

positive everywhere except the aforementioned point $\lambda_o = 0, \mu_o = 2m+1$. Therefore cubic polynomial (3.7) must have an odd (even) number of negative roots if ${}_1c_0 > 1$ (${}_1c_0 < 1$). In particular this implies that the polynomial has an odd number of positive roots if ${}_1c_0 < 1$ so that the AEH solution **a**′m always exists outside Area $A_m$ in this case.

Since two roots of the cubic polynomial in question cannot merge in Area $A_m$ we conclude that this is also true for any $\lambda_o$ between 0 and $\mu_o - 2m - 1 > 0$. In particular this implies the polynomial $G^{(m)}_3[\lambda\,|\,{}_1\boldsymbol{G}^{210}]$ has two negative roots (AEH solutions **a**m and **d**m) within Area $A_m$ if ${}_1c_0 > 1$ and a single negative root (AEH solution **d**m) if ${}_1c_0 < 1$.

At first glance the assertion that only the AEH solution **c**m may change its type along the **c**/**a′** separatrix contradicts the results of the previous Section. However the proof presented in [2] was necessarily based on the assumption that the RI $\lambda_o$ has a nonzero value and therefore is inapplicable to the AL limit of the *r*-GRef potential.

Since the regular AEH solution may not change its type as the RI $\lambda_o$ monotonically increases within Area $A_m$ at fixed values of $\mu_o$ and $z_T$ it must remain below the ground energy level if this is true for the appropriate AL potential curve. In particular this implies that AEH solutions **a**m ($_1c_0$ >1) are nodeless for m = 0, …, $m_{\mathbf{a};\mathbf{c}0}$, where

$$m_{\mathbf{a};\mathbf{c}0} = min\{[{}^{+}m_{\mathbf{c}0}(\mu_o;{}_1c_0 > 1)],[\tfrac{1}{2}(\mu_o - \lambda_o - 1)]\}, \qquad (3.11a)$$

with ${}^{+}m_{\mathbf{c}0}(\mu_o;{}_1c_0)$ standing for the positive zero of quadratic polynomial (2.31) which is independent of $\lambda_o$ by definition. It should be stressed that (3.9a) defines only the sufficient condition for the AEH solution **a**m to lie below the ground energy level. As mentioned above this nodeless solution also exists on the $D_m$-side of the border line $A_m|D_m$.

The upper bound $m_{\mathbf{b};\mathbf{c}0}$ for index m counting nodeless AEH solutions **b**m for $_1c_0$ < 1 is limited from below by even a more complicated condition

$$m_{\mathbf{b};\mathbf{c}0} = min\{[{}^{+}m_{\mathbf{c}0}(\mu_o;{}_1c_0 < 1)],[\tfrac{1}{2}(\mu_o + \lambda_o - 1)]\} \qquad (3.11b)$$

under the constraint

$$\lambda_o < \mu_o + 2[{}^{+}m_{\mathbf{c}0}(\mu_o;{}_1c_0 < 1)] + 1. \qquad (3.11b^*)$$

These restrictions directly follows from the fact that each of three AEH solutions has a distinct type along the ZFE separatrix $A_m|D_m$ for $_1c_0$ < 1 and as a result a pair of roots of the cubic polynomial $G_3^{(m)}[\lambda \mid {}_1\boldsymbol{G}^{210}]$ cannot merge inside Area $D_m$ (similarly to the assertion made by us for Area $A_m$). Therefore the AEH solution **b**m from the primary sequence exits at any point of Area $D_m$.

As pointed to in [2] the quartic polynomial $G_4^{(m)}[\lambda \mid {}_1\boldsymbol{G}]$ has the double root ${}_1\lambda_{1;\mathbf{t}m} = -m - 1$ along the **a**/**d**-DRt hyperbola so that interrelation formula (3.5) between the signed exponent differences ${}_1\lambda_{0;\dagger m}$ and ${}_1\lambda_{1;\dagger m}$ becomes ambiguous. An analysis of DRtTP reduction (2.35) of the general expression for the **a**/**d**-DRt hyperbola shows that

$$\mu_o^2 = \lambda_o^2 + (2m+1)^2 - 2(1-\sqrt{{}_1c_0})(2m+1)^2 < (\lambda_o + 2m+1)^2 \quad \text{for } {}_1c_0 < 1. \tag{3.12}$$

This implies that the cubic polynomial ${}_1G_3^{(m)}[\lambda;\lambda_o,\mu_o;{}_1c_0<1]$ has three distinct real roots (and thereby a positive discriminant ${}_1\Delta_G^{(m)}$) at any point of Area $A_m$ so that its roots are determined by the following formulas

$${}_1\lambda_{1;\dagger_k m} = -\frac{1}{3\,{}_1G_{3;3}^{(m)}}\left[{}_1G_{3;2}^{(m)} + 2Re(u_k C)\right] \quad \text{(k=1, 2, 3)}, \tag{3.13}$$

where $u_k$ are three roots of –1 and the parameter C is defined as an arbitrarily chosen cubic root of the complex number

$$C^3 \equiv [{}_1G_{3;2}^{(m)}]^3 - {}^9\!/\!_2\,{}_1G_{3;3}^{(m)}\,{}_1G_{3;2}^{(m)}\,{}_1G_{3;1}^{(m)} + {}^{27}\!/\!_2[{}_1G_{3;3}^{(m)}]^2\,{}_1G_{3;0}^{(m)} + {}^1\!/\!_2\, i\sqrt{{}_1\Delta_G^{(m)}}, \tag{3.14}$$

with the arguments of all the coefficients dropped for simplicity.

Since the **a**/**d**-DRt hyperbola crosses Area $A_m$ if ${}_1c_0 > 1$ it seems preferable to avoid this ambiguity by computing the discrete energy spectrum of potential (3.1) and two other sets of factorization energies in Area $A_m$ based on three distinct real roots ${}_1\lambda_{0;\dagger m}$ of the companion cubic polynomial ${}_1\tilde{G}_3^{(m)}[\lambda;\lambda_o,\mu_o;{}_1c_0>1]$. The crucial advantage of this alternative prescription comes from the fact that the fraction

$${}_1\lambda_{1;\dagger m} = \frac{\mu_o^2 - ({}_1\lambda_{0;\dagger m} + 2m+1)^2 + (1 - 2/\sqrt{{}_1c_0})({}_1\lambda_{0;\dagger m}^2 - \lambda_o^2)}{2({}_1\lambda_{0;\dagger m} + 2m+1)} \tag{3.15}$$

determining the second signed ExpDiff ${}_1\lambda_{1;\dagger m}$ does not have any ambiguities in Area $A_m$ for ${}_1c_0 > 1$. In fact an analysis of the equation

$$\tilde{G}_3^{(m)}[-2m-1 \mid {}_1\boldsymbol{G}^{210}] \equiv \{(2-\sqrt{{}_1c_0})[(2m+1)^2-\lambda_o^2]-\sqrt{{}_1c_0}\mu_o^2\}^2=0 \tag{3.16}$$

shows that the cubic polynomial in question may have the root ${}_1\lambda_{0;\dagger m}=-m-1$ iff

$$\mu_o^2=(2/\sqrt{{}_1c_0}-1)[(2m+1)^2-\lambda_o^2]. \tag{3.17}$$

Since the value of the RI $\mu_o$ is smaller than $\lambda_o+2m+1$ for ${}_1c_0=1$ and then monotonically decreases as ${}_1c_0$ increases this that the above assertion that fraction (3.15) has no ambiguity in Area $A_m$ as far as ${}_1c_0 > 1$. We can thus compute three distinct real roots of the cubic polynomial ${}_1\tilde{G}_3^{(m)}[\lambda;\lambda_o,\mu_o;{}_1c_0>1]$ in Area $A_m$ via the formula

$${}_1\lambda_{0;\dagger m}=-\frac{1}{3\,{}_1\tilde{G}_{3;3}^{(m)}}\left[{}_1\tilde{G}_{3;3}^{(m)}+2Re(u_k\tilde{C})\right] \quad (k=1, 2, 3), \tag{3.18}$$

where cube of the complex number $\tilde{C}$ is defined via (3.12) with ${}_1G_{3;j}^{(m)}$ and discriminant ${}_1\Delta_G^{(m)}$ changed for the coefficients ${}_1\tilde{G}_{3;j}^{(m)}$ and discriminant ${}_1\tilde{\Delta}_G^{(m)}$ of the cubic polynomial ${}_1\tilde{G}_3^{(m)}[\lambda;\lambda_o,\mu_o;{}_1c_0]$.

One of the remarkable features of discriminants ${}_1\Delta_G^{(m)}$ and ${}_1\tilde{\Delta}_G^{(m)}$ is that they always have the same sign. Indeed if one of the cubic polynomials ${}_1G_3^{(m)}[\lambda;\lambda_o,\mu_o;{}_1c_0]$ and ${}_1\tilde{G}_3^{(m)}[\lambda;\lambda_o,\mu_o;{}_1c_0]$ has only real roots than the roots of the second polynomial must be all real since they can be computed either via (3.5) or via (3.15) accordingly.

Let us now show that the RCSLE with the DRtTP Bose invariant

$$I^o[z;\varepsilon \mid {}_1\boldsymbol{G}^{210}]=\frac{1-\lambda_o^2}{4z^2}-\frac{1}{4(z-1)^2}+\frac{{}_1O_0^o}{4z\ (1-z)}+\varepsilon\,{}_1\wp[z;T_2] \tag{3.19}$$

where

$$ {}_1\wp[z;T_2] \equiv \frac{(z - z_T)^2}{4z^2(1-z)^2}, \qquad (3.19') $$

has two infinite sequence of nodeless regular AEH solutions regardless of the value of the RI $\lambda_o$. Substituting

$$ {}_1\lambda_{r;\mathbf{t}\mathrm{m}} = (\mathrm{m} + \tfrac{1}{2})\tau_{r;\mathbf{t}\mathrm{m}} \quad (r = 0,1) \qquad (3.20) $$

into both cubic equation (3.2) and fractional formula (3.5), making m tend to +∞, and setting

$$ \tau_{r;\mathbf{t}} \equiv \lim_{\mathrm{m}\to\infty} \tau_{r;\mathbf{t}\mathrm{m}} \quad (r = 0,1) \qquad (3.20') $$

leads to the cubic equation

$$ 8\sqrt{c_0}(1 - \sqrt{c_0})\tau^3_{1;\mathbf{t}} + 4(1 + \sqrt{c_0} - {}_1c_0)\tau^2_{1;\mathbf{t}} + 4\tau_{1;\mathbf{t}} + 1 = 0 \qquad (3.21) $$

and the fractional formula

$$ \tau_{0;\mathbf{t}} = \frac{-2\sqrt{{}_1c_0}\,\tau^2_{1;\mathbf{t}} - 2\tau_{1;\mathbf{t}} - 1}{2(\tau_{1;\mathbf{t}} + 1)}. \qquad (3.21^*) $$

with the coefficients independent of both RIs $\lambda_o$ and $\mu_o$. Since the infinite tails of these sequences necessarily lie in Area $B_m$ the AEH solutions from these tails belong to the primary sequence **d**m and to two secondary sequences **d**′m and **a**′m for ${}_1c_0 < 1$ or **b**′m for ${}_1c_0 > 1$. As a results three roots of cubic equation (3.15) can be directly obtained from (2.15), (2.22), and (2.22d) in the limit m→ ∞:

$$ \tau_{1;\breve{\mathbf{t}}_-} = \frac{1}{2(\sqrt{{}_1c_0} - 1)}, \qquad (3.22) $$

$\breve{\mathbf{t}}_- = \mathbf{a}'$ or $\mathbf{b}'$ for ${}_1c_0 < 1$ or ${}_1c_0 < 1$, respectively,

$$\tau_{1;\mathbf{d}} = \begin{cases} \dfrac{|\sqrt{{}_1c_0}-1|-\sqrt{{}_1c_0}-1}{4\sqrt{{}_1c_0}} & \text{for } {}_1c_0<1 \\ -\dfrac{\sqrt{{}_1c_0}+1+|\sqrt{{}_1c_0}-1|}{4\sqrt{{}_1c_0}} & \text{for } {}_1c_0<1 \end{cases}, \tag{3.22d′}$$

and

$$\tau_{1;\mathbf{d}''} = -\frac{\sqrt{{}_1c_0}+1+|\sqrt{{}_1c_0}-1|}{4\sqrt{{}_1c_0}} \tag{3.22d″}$$

taking into account that

$$\lim_{m\to\infty} \frac{{}_1\breve{\Delta}_m(\mu_o;{}_1c_0)}{(2m+1)^2} = 4(\sqrt{{}_1c_0}-1)^2. \tag{3.23}$$

Despite the fact that the AEH solutions $\mathbf{d}''m$ at the energies

$${}_1\varepsilon_{\mathbf{d}''m} \approx -\tau^2_{\mathbf{d}''m}(2m+1)^2$$

form a tail of the primary sequence $\mathbf{d}m$ in the limiting case of the WL potential we prefer to distinguish between these two sequences until it is proven that the primary sequence $\mathbf{d}m$ (m=0, 1, 2,…) does contain an infinite number of AEH solutions at least within the range $0 \le \lambda_o < \mu_o - 1$. By representing (3.20) as

$$2\tau_{1;\breve{\mathbf{t}}_-} + 1 = 2\sqrt{{}_1c_0}\,\tau_{1;\breve{\mathbf{t}}_-} \tag{3.22}$$

one can directly verify that $\tau_{0;\breve{\mathbf{t}}_-} = -\sqrt{{}_1c_0}\,\tau_{1;\breve{\mathbf{t}}_-}$ as expected. Keeping in mind $\tau_{1;\mathbf{d}'}$ and $\tau_{1;\mathbf{d}''}$ introduced via (3.20d) and (3.20d′) are nothing two roots of the quadratic equation

$$2\sqrt{{}_1c_0}\,(2\,{}_1\tau_{1;\breve{\mathbf{t}}_+} + 1)\,{}_1\tau_{1;\breve{\mathbf{t}}_+} = -2\,{}_1\tau_{1;\breve{\mathbf{t}}_+} - 1 \tag{3.22d}$$

we also confirm that $\tau_{0;\breve{\mathbf{t}}_+} = \sqrt{{}_1c_0}\,\tau_{1;\breve{\mathbf{t}}_+}$.

We thus conclude that the RCSLE with the DRtTP Bose invariant has infinitely many regular AEH solutions which necessarily lie below the ground energy level at sufficiently large m since $|\,{}_1\varepsilon_{\breve{\mathbf{t}}\text{-}m}\,|$ grows as $m^2$. These AEH solutions can be therefore used as seed functions to construct infinite sequences of rational potentials conditionally exactly quantized by GS Heine polynomials.

## 4. Quantization of the *e*-DRtTP potential in terms of elementary functions

It has been shown by Roychoudhury et al [14] that first-order differential equation (1.3) has the solution

$$\hat{\eta}(x) = \sqrt{\exp(2\sqrt{C}x) + 1} \tag{4.1}$$

which can be converted into the variable of our interest

$$z(x) = \frac{2}{1 + \sqrt{1 + e^{-2x}}} \tag{4.2}$$

via the linear-fractional transformation

$$z(x) = \frac{2}{\hat{\eta}(x) + 1}, \tag{4.3}$$

with $\sqrt{C}$ set to −1. As it has been pointed in Introduction variable (4.2) satisfies the differential equation

$$z' = \frac{2z(1-z)}{2-z} \tag{4.4}$$

and therefore belongs to the class of the transformation converting the Schrödinger equation with DRtTP potential (3.1) into the hypergeometric equation.

To compare our general representation (3.1) for the DRtTP potential $V[z(x)\,|\,{}_1\boldsymbol{G}^{210}(z_T)]$ with the DKV potential[13] first remember that the variables $\hat{\eta}$ and z are related via the linear fractional transformation and therefore the Schwartz derivative $\{\hat{\eta}, z\}$ vanishes so that the Schwartz derivatives $\{\hat{\eta}, x\}$ and $\{z, x\}$ must coincide [17]. Substituting

$$2 - z = \frac{2\hat{\eta}}{\hat{\eta}+1}, \tag{4.5}$$

together with (4.3) and a similar relation for 1–z, into (2.10) thus gives

$$\{\hat{\eta}, x\} = -\tfrac{1}{2} - 3\hat{\eta}^{-2} + \tfrac{3}{2}\hat{\eta}^{-4} . \tag{4.6}$$

The very remarkable feature of variable (4.1) is that its Schwartz derivative contains only even powers of $\hat{\eta}^{-1}$. As a result the expression for potential (3.1) in terms of this variable does not contain the term proportional to $\hat{\eta}^{-3}$.

Since $\{\hat{\eta}, z\} = 0$ the RefPFs associated with the variables $\hat{\eta}$ and z are interrelated in a simple fashion [17]

$$\hat{I}^{o}[\hat{\eta}; \lambda_o, \mu_o] = \left(\frac{dz}{d\hat{\eta}}\right)^2 {}_1I^{o}[2/(\hat{\eta}+1); \lambda_o, \mu_o]. \tag{4.7}$$

Substituting

$$\left(\frac{dz}{d\hat{\eta}}\right) = -\frac{2}{(\hat{\eta}+1)^2} \tag{4.8}$$

and (2.11) into the right-hand side of (4.7) thus gives

$$\hat{I}^{o}[\hat{\eta}; \lambda_o, \mu_o] = \frac{1-\mu_0^2}{4(\hat{\eta}+1)^2} + \frac{1}{4(1-\hat{\eta})^2} + \frac{\mu_o^2 - \lambda_o^2}{4(\hat{\eta}^2 - 1)} . \tag{4.9}$$

Note that the ExpDiff at the singular point $\hat{\eta} = -1$ is equal to $\mu_o$ as expected from the fact that linear fractional transformation (4.3) converts this singular point onto infinity. Substituting and Schwartz derivative (4.6), together with the first derivative of $\hat{\eta}$ with respect to x,

$$\hat{\eta}' = \frac{(1-\hat{\eta}^2)}{\hat{\eta}} \tag{4.10}$$

into the general expression for the Liouville potential thus gives

$$V[2/(\hat{\eta}+1)\,|\,{}_1\boldsymbol{G}^{210}(2)] = \tfrac{1}{4}\lambda_o^2 - \frac{\mu_0^2}{2\hat{\eta}} + \frac{2\mu_0^2 - \lambda_o^2 + 3}{4\hat{\eta}^2} - \frac{3}{4\hat{\eta}^4}. \tag{4.11}$$

The potential differs from the DKV potential [13]

$$V_{DKV}[\hat{\eta};A,B] \equiv -B\hat{\eta}^{-1} + A\hat{\eta}^{-2} - \tfrac{3}{4}\hat{\eta}^{-4} \tag{4.12}$$

by a shift of the zero point energy, namely,

$$V_{DKV}[\hat{\eta};A,B] - V_{DKV}[1;A,B] = V[2/(\hat{\eta}+1)\,|\,{}_1\boldsymbol{G}^{210}(2)](\hat{\eta}-1), \tag{4.13}$$

where

$$V_{DKV}[1;A,B] = A - B - \tfrac{3}{4}. \tag{4.14}$$

Comparing (4.12) with (4.11) we can thus come to the following formulas for the DKV parameters A and B:

$$A = \tfrac{1}{4}(2\mu_o^2 - \lambda_o^2 + 3), \tag{4.15a}$$

$$B = \tfrac{1}{2}\mu_o^2, \tag{4.15b}$$

and

$$V_{DKV}[1;A,B] = -\tfrac{1}{4}\lambda_o^2. \tag{4.16}$$

The energy-dependent parameters $\alpha$ and $\beta$ in [14] are nothing but the signed ExpDiffs for the singular points $\hat{\eta} = \pm 1$:

$$\alpha = {}_1\lambda_{1;\mathbf{c}n} \text{ and } \beta = -\mu_{\mathbf{c}n}. \tag{4.17}$$

To verify (4.16) one simply needs to compare the identity

$$z^{\frac{1}{2}({}_1\lambda_{0;\mathbf{c}v}+2m+1)}(1-z)^{\frac{1}{2}({}_1\lambda_{1;\mathbf{c}v}+1)}$$
$$=2^{\frac{1}{2}({}_1\lambda_{0;\mathbf{c}v}+2m+1)}(\hat{\eta}-1)^{\frac{1}{2}({}_1\lambda_{1;\mathbf{c}v}+1)}(\hat{\eta}+1)^{-\frac{1}{2}(\mu_{\mathbf{c}v}+1)}, \tag{4.18}$$

with Eq. (3.14) in [14] for the eigenfunctions of the Schrödinger equation with the DKV potential keeping in mind that the latter differ from the eigenfunctions of the appropriate RCSLE by the quartic root of the weight function

$$\hat{\wp}[\hat{\eta}]=\frac{\hat{\eta}^2}{(1-\hat{\eta}^2)^2}. \tag{4.19}$$

By expressing the Jacobi polynomial in $\hat{\eta}\equiv 2\hat{z}-1$ as the hypergeometric polynomial in $\hat{z}\equiv 1/z$ we come to the generic expression [1] for the hypergeometric polynomials describing eigenfunctions for the *r*-GRef potential:

$$F[-n,\ \mu_{\mathbf{c}n}-n;\ {}_1\lambda_{0;\mathbf{c}n}+1;z]=(-\hat{z})^{-n}\frac{\Gamma(-\lambda_{1;\mathbf{c}n}-n)\Gamma(n+1-\mu_{\mathbf{c}n})}{\Gamma(1-\mu_{\mathbf{c}n})\Gamma(-\lambda_{1;\mathbf{c}n})}$$
$$\times F[-n,{}_1\lambda_{1;\mathbf{c}n}-\mu_{\mathbf{c}n}+n+1;1-\mu_{\mathbf{c}n};\hat{z}]. \tag{4.20}$$

Since ${}_1a_2=1$ and ${}_1c_0=4$ for $z_T=2$ the squares of the ExpDiffs $|\,{}_1\lambda_{0;\mathbf{t}m}|$, $|\,{}_1\lambda_{1;\mathbf{t}m}|$, and $|\,\mu_{\mathbf{t}m}|$ are related as follows

$$\mu^2_{\mathbf{c}n}-{}_1\lambda^2_{1;\mathbf{c}n}=\mu^2_o. \tag{4.21}$$

and

$${}_1\lambda^2_{0;\mathbf{c}n}=\lambda^2_o+4\,{}_1\lambda^2_{1;\mathbf{c}n}. \tag{4.21*}$$

The first formula is nothing but (3.10) in [14] whereas the second allows us to represent the eigenvalues $E_n$ given by (3.13) in [14] as

$$E_n=-\tfrac{1}{4}\,{}_1\lambda^2_{0;\mathbf{c}n} \tag{4.22a}$$

$$=-\tfrac{1}{4}\lambda^2_o+{}_1\varepsilon_{\mathbf{c}n}, \tag{4.22b}$$

where ${}_1\lambda_{0;\mathbf{c}n}$ is a positive root of cubic polynomial (3.8) such that

$$\frac{\mu_o^2-({}_1\lambda_{0;\mathbf{c}n}+2n+1)^2+(1-2/\sqrt{{}_1c_0})({}_1\lambda_{0;\mathbf{c}n}^2-\lambda_o^2)}{2({}_1\lambda_{0;\mathbf{c}n}+2n+1)}>0. \tag{4.23}$$

The constant energy difference is caused by shift (4.15*) of the zero-point energy for the DKV potential compared with (4.12). Finally, representing (3.11) in [14] as

$$A=-\tfrac{1}{4}\,{}_1\lambda_{0;\mathbf{c}n}^2+\tfrac{1}{4}(\mu_{\mathbf{c}n}-{}_1\lambda_{1;\mathbf{c}n})^2+\tfrac{1}{4}(\mu_{\mathbf{c}n}+{}_1\lambda_{1;\mathbf{c}n})^2+\tfrac{3}{4} \tag{4.24}$$

$$=-\tfrac{1}{4}({}_1\lambda_{0;\mathbf{c}n}^2-4\,{}_1\lambda_{1;\mathbf{c}n}^2)+\tfrac{1}{2}(\mu_{\mathbf{c}n}^2-{}_1\lambda_{1;\mathbf{c}n}^2)+\tfrac{3}{4} \tag{4.24*}$$

and again making use of (4.21) and (4.21*) we confirm that definition (3.11) of the parameter A in [14] agrees with (4.15a).

## 5. Use of basic solutions for constructing a triad of DRtTP potentials quantized by Heun polynomials

The benchmark feature of the generic DRtTP *r*-GRef potential is that its three ($\mathbf{t}=\breve{\mathbf{t}}_{-}$, $\mathbf{c}$, or $\mathbf{d}$) SUSY partners $V[z\,|\,{}^{1}_{1}\boldsymbol{G}^{210}_{\mathbf{t}0}]$ are exactly quantized by Heun polynomials [2]. Indeed each CLDT using the basic solution

$$\phi_{\mathbf{t}0}[z\,|\,{}_1\boldsymbol{G}^{210}_{\downarrow\mathbf{t}0}]=z^{\,{}_1\rho_{0;\mathbf{t}0}}(1-z)^{\,{}_1\rho_{1;\mathbf{t}0}} \tag{5.1}$$

$$\equiv\sqrt{z(1-z)}\;z^{\frac{1}{2}\,{}_1\lambda_{0;\mathbf{t}0}}(1-z)^{\frac{1}{2}\,{}_1\lambda_{1;\mathbf{t}0}} \tag{5.1$'$}$$

as its FF transforms the PFB ${}_1\boldsymbol{G}^{210}_{\downarrow\mathbf{t}0}$ into another PFB ${}^{1}_{1}\boldsymbol{G}^{210}_{\mathbf{t}0}$ with the RefPF given by

$$I^{o}[z\,|\,{}^{1}_{1}\boldsymbol{G}^{210}_{\mathbf{t}0}(z_T)]=I^{o}[z\,|\,{}_1\boldsymbol{G}^{210}_{\downarrow\mathbf{t}0}(z_T)]-\frac{2}{(z-z_T)^2}+\frac{\Delta O_1[z\,|\,{}^{1}_{1}\boldsymbol{G}^{210}_{\mathbf{t}0}(z_T)]}{4z(z-1)(z-z_T)}, \tag{5.2}$$

where, according to (C.8*) in [2] for ι=1, $\mathfrak{I}$=1 and m = 0,

$$\Delta O_1[z\,|\,{}^{1}_{1}\boldsymbol{G}^{210}_{\dagger 0}(z_T)] = 4[2z-(\mu_{\dagger 0}-1)z_T + {}_1\lambda_{0;\dagger 0}+1] \tag{5.2′}$$

so that the resultant RCSLE has four regular singular points 0, 1, $z_T$, and $\infty$.

The associated RLP thus takes the form

$$V[z\,|\,{}^{1}_{1}\boldsymbol{G}^{210}_{\dagger 0}(z_T)] = V[z\,|\,{}_{1}\boldsymbol{G}^{210}_{\downarrow\dagger 0}(z_T)] + \frac{8z^2(z-1)^2}{(z-z_T)^4} - \frac{z(z-1)\Delta O_1[z\,|\,{}^{1}_{1}\boldsymbol{G}^{210}_{\dagger 0}]}{(z-z_T)^3}. \tag{5.3}$$

For $z_T = 2$ the potential is explicitly expressible in terms of x:

$$\begin{aligned} V[z(x)\,|\,{}^{1}_{1}\boldsymbol{G}^{210}_{\dagger 0}(2)] = V[z(x)\,|\,{}_{1}\boldsymbol{G}^{210}_{\downarrow\dagger 0}(2)] &+ \frac{8z^2(x)[z(x)-1]^2}{[z(x)-2]^4} \\ &- \frac{z(x)[z(x)-1]\Delta O_1[z(x)\,|\,{}^{1}_{1}\boldsymbol{G}^{210}_{\dagger 0}(2)]}{[z(x)-2]^3}, \end{aligned} \tag{5.3*}$$

where the change of variable z(x) is an elementary function defined via (4.2).

The energy-dependent gauge transformation,

$$\Phi[z;\varepsilon\,|\,{}^{1}_{1}\boldsymbol{G}^{210}_{\dagger 0}] = z^{\rho_0(\varepsilon|{}_1\boldsymbol{G}_{\downarrow\dagger 0};\bar{\sigma})}(1-z)^{\rho_1(\varepsilon|{}_1\boldsymbol{G}_{\downarrow\dagger 0};\bar{\sigma})}(z-z_T)^{-1}\mathrm{Hf}[z;\varepsilon\,|\,{}^{1}_{1}\boldsymbol{G}^{210}_{\dagger 0};\bar{\sigma}], \tag{5.4}$$

obtained from (3.10) in [2] by setting $\iota$, $\ell$ K, and $\Im$ to 1, 1, 2, and 1 accordingly, converts the RCSLE into the Heun equation

$$\begin{aligned} \Big\{z(z-1)(z-z_T)\frac{d^2}{dz^2} &+ 2B_2[z;\bar{\rho}(\varepsilon\,|\,{}_{1}\boldsymbol{G}^{210}_{\downarrow\dagger 0;\dagger' 0};\bar{\sigma});z_T]\frac{d}{dz} \\ &+ C_1[z;\varepsilon\,|\,{}^{1}_{1}\boldsymbol{G}^{210}_{\dagger 0};\bar{\sigma}]\Big\}\mathrm{Hf}[z;\varepsilon\,|\,{}^{1}_{1}\boldsymbol{G}^{210}_{\dagger 0};\bar{\sigma}] = 0. \end{aligned} \tag{5.5}$$

The second- and first-order polynomials in the left-hand side of this differential equation are defined as follows

$$\begin{aligned} B_2[z;\bar{\rho}(\varepsilon\,|\,{}_{1}\boldsymbol{G}^{210}_{\downarrow\dagger 0};\bar{\sigma});z_T] &\equiv z(z-1)(z-z_T) \\ &\times\Big[\frac{\rho_0(\varepsilon\,|\,{}_{1}\boldsymbol{G}^{210}_{\downarrow\dagger 0};\bar{\sigma})}{z} + \frac{\rho_1(\varepsilon\,|\,{}_{1}\boldsymbol{G}^{210}_{\downarrow\dagger 0};\bar{\sigma})}{z-1} - \frac{1}{z-z_T}\Big], \end{aligned} \tag{5.6}$$

where

$$\rho_r(\varepsilon\,|\,{}_1\mathbf{G};\bar{\sigma}) \equiv \begin{cases} \tfrac{1}{2}(\sigma_0\sqrt{\lambda_o^2 - {}_1c_0\,\varepsilon}+1) & \text{for } r=0, \\ \tfrac{1}{2}(\sigma_1\sqrt{-\varepsilon}+1) & \text{for } r=1, \end{cases} \tag{5.7}$$

and

$$C_1[z;\varepsilon\,|\,{}^{1}_{1}\mathbf{G}^{210}_{\dagger 0};\bar{\sigma}] = \alpha(\varepsilon\,|\,{}^{1}_{1}\mathbf{G}^{210}_{\dagger 0};\bar{\sigma})\beta(\varepsilon\,|\,{}^{1}_{1}\mathbf{G}^{210}_{\dagger 0};\bar{\sigma})z - q(\varepsilon\,|\,{}^{1}_{1}\mathbf{G}^{210}_{\dagger 0};\bar{\sigma}), \tag{5.8}$$

where the energy-dependent α- and β-coefficients are related to characteristic exponents (5.7) via the conventional formula [22]

$$\alpha(\varepsilon\,|\,{}^{1}_{1}\mathbf{G}^{210}_{\dagger 0};\bar{\sigma}) + \beta(\varepsilon\,|\,{}^{1}_{1}\mathbf{G}^{210}_{\dagger 0};\bar{\sigma}) = \rho_0(\varepsilon\,|\,{}_1\mathbf{G}^{210}_{\downarrow\dagger 0};\sigma_0) + \rho_1(\varepsilon\,|\,{}_1\mathbf{G}^{210}_{\downarrow\dagger 0};\sigma_1) - 3 \tag{5.9}$$

The energy-dependent accessory parameter is given by (3.29) and (3.30) in [2], with K=2, $\mathfrak{I}$=1, n = 0, $\ell$ = 1, namely,

$$q(\varepsilon\,|\,{}^{1}_{1}\mathbf{G}^{210}_{\dagger 0};\bar{\sigma}) = -C_{1;0}\,|\,\varepsilon\,|\,{}^{1}_{\iota}\mathbf{G}^{210}_{\dagger 0};\bar{\sigma}) = -\tfrac{1}{4}O_{1;0}\,|\,{}^{1}_{\iota}\mathbf{G}^{210}_{\dagger 0}) \tag{5.10}$$

$$+\tfrac{1}{4}\,{}_1d\,\varepsilon - 2\rho_0(\varepsilon\,|\,{}_1\mathbf{G}^{210}_{\downarrow\dagger 0};\bar{\sigma})\rho_1(\varepsilon\,|\,{}_1\mathbf{G}^{210}_{\downarrow\dagger 0};\bar{\sigma})z_T - 2\rho_0(\varepsilon\,|\,{}_1\mathbf{G}^{210}_{\downarrow\dagger 0};\bar{\sigma}).$$

When deriving (5.10) we also took into account that polynomials $\hat{O}^{\downarrow}_1[z\,|\,{}^{1}_{1}\mathbf{G}^{210}_{\dagger 0}]$ and $O_1[z\,|\,{}^{1}_{1}\mathbf{G}^{210}_{\dagger 0}]$ concur since the PFB ${}^{1}_{1}\mathbf{G}^{210}_{\dagger 0}$ does not have any other outer roots (n=0) so that PF (3.26) in [2] coincides with

$$\hat{Q}^{210}[z;z_T] = -\frac{1}{(z-z_T)^2}. \tag{5.11}$$

There are 8 gauge transformation which convert the given RCSLE into the Heun equation. Here we are only interested in three of them which have Gauss-seed Heun polynomials as their solutions, namely, either $\sigma_0 = \sigma_1 = \pm$, $\sigma_2 = -$ or $\sigma_0 = \sigma_2 = -$, $\sigma_1 = +$ and $\sigma_0 = +$, $\sigma_1 = \sigma_2 = -$ for ${}_1c_0 < 1$ and ${}_1c_0 > 1$, accordingly.

One can directly relate the energy-dependent parameters $\alpha(\varepsilon\,|\,{}^{1}_{1}\mathcal{G}^{210}_{\dagger 0};\bar{\sigma})$ and $\beta(\varepsilon\,|\,{}^{1}_{1}\mathcal{G}^{210}_{\dagger 0};\bar{\sigma})$ to their counter-parts $\alpha(\varepsilon\,|\,{}_{1}\mathcal{G}^{210};\bar{\sigma})$ and $\beta(\varepsilon\,|\,{}_{1}\mathcal{G}^{210};\bar{\sigma})$ in the hypergeometric equation for the *r*-GRef PFB ${}_{1}\mathcal{G}^{210}$ by taking advantage of the general theorem proven in [2] for SUSY pairs of RCSLE with regular singular points (including infinity) assuming that they are generated by means of a second-order TP. Namely, let

$$I^{o}[z\,|\,{}^{0}_{1}\mathbf{B}^{2\Im 0}] = \frac{1-\lambda_{o}^{2}}{4z^{2}} + \frac{1}{4(1-z)^{2}} - \sum_{k=1}^{n} \frac{2}{(z-z^{(n)}_{o;k})^{2}} + \frac{O^{o}_{n+1}[z\,|\,{}^{0}_{1}\mathbf{B}^{2\Im 0}]}{4z(z-1)\Pi_{n}[z;\bar{z}^{(n)}_{o}]} \tag{5.12}$$

be the RefPF of the RCSLE with n+3 regular singular points (including infinity). Note that all n outer singularities $z^{(n)}_{o;k}$ ($z^{(n)}_{o;k} < 0$ or $z^{(n)}_{o;k} > 1$) have by definition energy-independent ExpDiffs equal to 3 since the appropriate PFB is assumed to be obtained from by a p-step CLDT using p GS solutions $\dagger_{j}m_{j}$. This CLDT converts any other GS solution $\dagger m$ into the solution of the given RCSLE which has the AEH form:

$$\phi[z\,|\,{}^{0}_{1}\mathbf{B}^{2\Im 0}_{\downarrow\dagger m};\dagger m] \propto \sqrt{z(1-z)}\; z^{\frac{1}{2}{}_{1}\lambda_{0;\dagger m}} (1-z)^{\frac{1}{2}{}_{1}\lambda_{1;\dagger m}} \frac{\Pi_{n_{\dagger m}}[z;{}^{*}\bar{z}^{(n_{\dagger m})}_{\dagger m}]}{\Pi_{n}[z;\bar{z}^{(n)}_{o}]}. \tag{5.13}$$

We then use this solution as the FF to construct a P-CES SUSY partner ${}^{1}_{1}\mathbf{B}^{2\Im 0}_{\dagger 0}$ of the PFB ${}^{0}_{1}\mathbf{B}^{2\Im 0}$. Finally the RCSLEs associated with both PFBs are converted into the Heine differential equations [21, 22] with polynomial coefficients by using the appropriate gauge transformations. As proven in Section 8 in [2] the energy-dependent parameters $\alpha(\varepsilon\,|\,{}^{1}_{1}\mathbf{B}^{2\Im 0}_{\dagger m};\bar{\sigma})$ and $\beta(\varepsilon\,|\,{}^{1}_{1}\mathbf{B}^{2\Im 0}_{\dagger 0};\bar{\sigma})$ are related to their counter-parts $\alpha(\varepsilon\,|\,{}^{0}_{1}\mathbf{B}^{2\Im 0}_{\downarrow\dagger m};\bar{\sigma})$ and $\beta(\varepsilon\,|\,{}^{0}_{1}\mathbf{B}^{2\Im 0}_{\downarrow\dagger m};\bar{\sigma})$ in a simple fashion:

$$\alpha(\varepsilon\,|\,{}^{1}_{1}\mathbf{B}^{2\Im 0}_{\dagger m};\bar{\sigma}) = \alpha(\varepsilon\,|\,{}^{0}_{1}\mathbf{B}^{2\Im 0}_{\downarrow\dagger m};\bar{\sigma}) + n - n_{\dagger m} - 1 \tag{5.14a}$$

and

$$\beta(\varepsilon \mid {}^{1}_{1}\mathbf{B}^{2\Im 0}_{\dagger m};\bar{\sigma}) = \beta(\varepsilon \mid {}^{0}_{1}\mathbf{B}^{2\Im 0}_{\downarrow\dagger m};\bar{\sigma}) + n - n_{\dagger m} - 1. \qquad (5.14b)$$

In case of our current interest $n = n_{\dagger m} = 0$ so that

$$\alpha(\varepsilon \mid {}^{1}_{1}\boldsymbol{G}^{210}_{\dagger 0};\bar{\sigma}) = \alpha(\varepsilon \mid {}_{1}\boldsymbol{G}^{210}_{\downarrow\dagger 0};\bar{\sigma}) - 1 \qquad (5.15a)$$

and

$$\beta(\varepsilon \mid {}^{1}_{1}\boldsymbol{G}^{210}_{\dagger 0};\bar{\sigma}) = \beta(\varepsilon \mid {}_{1}\boldsymbol{G}^{210}_{\downarrow\dagger 0};\bar{\sigma}) - 1. \qquad (5.15b)$$

Let us now consider another GS solution $\dagger' m'$ at the energy ${}_{1}\varepsilon_{\dagger' m'}$ so that either

$$\alpha({}_{1}\varepsilon_{\dagger' m'} \mid {}_{1}\boldsymbol{G}^{210}_{\downarrow\dagger 0};\bar{\sigma}_{\dagger'}) = -m' \qquad (5.16a)$$

or

$$\beta({}_{1}\varepsilon_{\dagger' m'} \mid {}_{1}\boldsymbol{G}^{210}_{\downarrow\dagger 0};\bar{\sigma}_{\dagger'}) = -m' \qquad (5.16b)$$

depending on the solution ($\alpha$- or $\beta$-) Class. Making use of (5.15a) or (5.15b), accordingly, thus shows that the order of the partner Heun polynomial is equal to

$$n_{\dagger' m'} = m' + 1. \qquad (5.17)$$

Altogether there are 9 primary sequences $\dagger 0;\dagger' m'$ of Heun polynomials

$$\mathrm{Hp}_{m'+1}[z \mid {}^{1}_{1}\boldsymbol{G}^{210}_{\dagger 0\downarrow\dagger' m'};\dagger' m'] \propto {}_{1}\hat{g}({}_{1}\lambda_{0;\dagger' m'} - {}_{1}\lambda_{0;\dagger 0}, {}_{1}\lambda_{1;\dagger' m'} - {}_{1}\lambda_{1;\dagger 0})\Pi_{m'}[\xi;\bar{\xi}_{\dagger' m'}] \qquad (5.18)$$

where the 'Heine-polynomial generator' (HPG) is defined via (7.10*):

$${}_{\iota}\hat{g}(\Delta\lambda, \Delta\nu) \equiv -\prod_{r=0}^{|\iota|} (\xi - {}_{\iota}e_{r})\frac{d}{d\xi} - {}_{\iota}\hat{B}_{1}[\xi;\bar{\iota};\Delta\lambda,\Delta\nu]. \qquad (5.19)$$

Any polynomial in each sequence is explicitly expressible in terms of hypergeometric polynomials as follows:

$$\frac{(-1)^{m'}(\mu_{\mathbf{t}0}-\mu_{\mathbf{t}'m'})(\mu_{\mathbf{t}'m'}-m')_{m'}}{({}_1\lambda_{0;\mathbf{t}'m'}+2)_{m'-1}}\,\mathrm{Hi}_{m'+1}[z\,|\,{}^{1}_{1}\boldsymbol{G}^{210}_{\mathbf{t}0\downarrow\mathbf{t}'m'};\mathbf{t}'m'] \tag{5.20}$$

$$\begin{aligned}
&= m'({}_1\lambda_{0;\mathbf{t}'m'}+{}_1\lambda_{1;\mathbf{t}'m'}+m'+1)z(z-1)\\
&\qquad\qquad\times F(1-m',\mu_{\mathbf{t}'m'}-m'+1;{}_1\lambda_{0;\mathbf{t}'m'}+2;z)\\
&+\tfrac{1}{2}({}_1\lambda_{0;\mathbf{t}'m'}+1)[({}_1\lambda_{0;\mathbf{t}'m'}-{}_1\lambda_{0;\mathbf{t}0})(z-1)+({}_1\lambda_{1;\mathbf{t}'m'}-{}_1\lambda_{1;\mathbf{t}0})z]\\
&\qquad\qquad\times F(-m',\mu_{\mathbf{t}'m'}-m';{}_1\lambda_{0;\mathbf{t}'m'}+1;z),
\end{aligned}$$

where

$$\mu_{\mathbf{t}'m'} \equiv {}_1\lambda_{0;\mathbf{t}'m'}+{}_1\lambda_{1;\mathbf{t}'m'}+2m'+1. \tag{5.21}$$

Two pairs of these sequences, $\mathbf{t}0;\tilde{\mathbf{t}}m'$ and $\tilde{\mathbf{t}}0;\mathbf{t}m'$ ($\mathbf{t}=\breve{\mathbf{t}}_-$, $\tilde{\mathbf{t}}=\mathbf{c}$ or $\mathbf{d}$ for $m'=0,1,\ldots$) start from the common first-order polynomial $z-z_{\mathbf{t}\tilde{\mathbf{t}}}$. Similarly the polynomial $z-z_{\mathbf{cd}}$ serves as the first term in the sequence $\mathbf{c}0;\mathbf{d}m$ (m = 0, 1,…). On other hand, since the ground energy eigenfunction for the RLP $V[z\,|\,{}^{1}_{1}\boldsymbol{G}^{210}_{\mathbf{d}0}]$ is given by the AEH solution ${}^{*}\phi_{\mathbf{d}0}[z\,|\,{}_1\boldsymbol{G}^{210}_{\downarrow\mathbf{d}0}]$ one also need to add a constant to the sequence $\mathbf{c}0;\mathbf{d}m$ (m = 0, 1,…). Finally three other primary sequences $\mathbf{t}0;\mathbf{t}m$ ($\mathbf{t}=\breve{\mathbf{t}}_-$, $\mathbf{c}$, or $\mathbf{d}$; $m=1,2,\ldots$) start from the second-order polynomial $\mathrm{Hp}_2[z\,|\,{}^{1}_{1}\boldsymbol{G}^{210}_{\mathbf{t}0\downarrow\mathbf{t}1};\mathbf{t}1]$.

Keeping in mind that the GS Heun polynomials forming AEH solutions $\breve{\mathbf{t}}_-0;\breve{\mathbf{t}}_-m$ (m > 0) and $\mathbf{c}0;\breve{\mathbf{t}}_-m$ (m = 0, 1,…) below the energies ${}_1\varepsilon_{\mathbf{c}0}$ and ${}_1\varepsilon_{\mathbf{c}1}$ accordingly have zeros only outside the quantization interval [0, 1] we can use these solutions as FFs to construct new RLPs conditionally exactly quantized by GS Heine polynomials. In addition there is an infinite sub-set of a secondary sequence of Heun polynomials $\mathrm{Hp}_m[z\,|\,{}^{1}_{1}\boldsymbol{G}^{210}_{\mathbf{t}0\downarrow\hat{\mathbf{t}}'_-};\hat{\mathbf{t}}'_-]$ with all the zeros outside the quantization interval [0, 1]. [Note that the type of AEH solutions $\breve{\mathbf{t}}'_-m$ from the secondary sequence (i.e., $\mathbf{b}$ if $c_0>1$ and $\mathbf{a}$ if $c_0<1$) differs from the type of AEH solutions $\breve{\mathbf{t}}_-m$ from the primary sequence ($\breve{\mathbf{t}}_-=\mathbf{a}$ or $\mathbf{b}$ if $c_0>1$ or $c_0<1$ respectively).] The general expression for the RefPFs $I^{o}[z\,|\,{}_{1}^{2}\boldsymbol{G}^{2\Im 0}_{\dagger 0;\dagger' m}]$ has been already obtained in [2] and is given by (6.30b) and (6.30b′) there.

Let us emphasize once again all the SUSY partners of the DKV potential ($z_T = 2$) are explicitly expressible in terms of x via the change of variable (4.2) and can be therefore quantized by elementary functions. In this regard we see no fundamental difference between these ‘explicit’ potentials and multi-step rational SUSY partners of the Rosen-Morse potential [6] constructed in [12, 26]

Note that, in addition to polynomial solutions (5.18), Heun equation (5.5) has Lambe and Ward’s ‘quasi-algebraic’ [27] solutions

$$f[z\,|\,{}_{1}^{1}\boldsymbol{G}^{210}_{\dagger 0};\dagger' m';\bar{\sigma}] = z^{\frac{1}{2}({}_{1}\lambda_{0;\dagger' m'} - \sigma_0 |{}_{1}\lambda_{0;\dagger' m'}|}(1-z)^{\frac{1}{2}({}_{1}\lambda_{1;\dagger' m'} - \sigma_1 |{}_{1}\lambda_{1;\dagger' m'}|} \times \mathrm{Hi}_{m'+1}[z\,|\,{}_{1}^{1}\boldsymbol{G}^{210}_{\dagger 0\downarrow\dagger' m'};\dagger' m'] \qquad (5.22)$$

which are also called ‘Heun polynomials’ in [19]. However we prefer to preserve the latter term solely for polynomial solutions (5.17) while referring to (5.19), in following [28], as the ‘Lambe-Ward kernels’.

## 6. Double-step SUSY partners conditionally exactly quantized by Heun polynomials

The CLDT using basic solution (5.1) as its FF converts another basic solution $\dagger' 0$ of type $\dagger'$ into the 1st-order AEH solution of partner RST equation for the PFB ${}_{1}^{1}\boldsymbol{G}^{210}_{\dagger 0\downarrow\dagger' 0}(z_T)$:

$$\phi[z\,|\,{}_{1}^{1}\boldsymbol{G}^{210}_{\dagger 0\downarrow\dagger' 0}(z_T);\dagger' 0] \propto \frac{W\{{}_{1}\phi_{\dagger 0}[z],{}_{1}\phi_{\dagger' 0}[z]\}}{{}_{1}\wp^{\frac{1}{2}}[z;{}_{1}T_2]\,{}_{1}\phi_{\dagger 0}[z]} \qquad (6.1)$$

$$= (\mu_{\dagger' 0} - \mu_{\dagger 0})\sqrt{z(1-z)}\, z^{\frac{1}{2}{}_{1}\lambda_{0;\dagger' 0}}\,(1-z)^{\frac{1}{2}{}_{1}\lambda_{1;\dagger' 0}}\,\frac{z - {}_{1}z_{\dagger\dagger'}}{z - z_T}, \qquad (6.1^*)$$

where

$$z_{\dagger\dagger'} = \frac{{}_{1}\lambda_{0;\dagger' 0} - {}_{1}\lambda_{0;\dagger 0}}{{}_{1}\lambda_{0;\dagger' 0} + {}_{1}\lambda_{1;\dagger' 0} - {}_{1}\lambda_{0;\dagger 0} - {}_{1}\lambda_{1;\dagger 0}} \qquad (6.2)$$

$$= \frac{{}_1\lambda_{0;\mathbf{t}'0} - {}_1\lambda_{0;\mathbf{t}0}}{\mu_{\mathbf{t}'0} - \mu_{\mathbf{t}0}} \tag{6.2a}$$

or, which is equivalent,

$$1 - z_{\mathbf{tt}'} == \frac{{}_1\lambda_{1;\mathbf{t}'0} - {}_1\lambda_{1;\mathbf{t}0}}{\mu_{\mathbf{t}'0} - \mu_{\mathbf{t}0}}. \tag{6.2b}$$

An analysis of the reciprocal

$$z_{\mathbf{tt}'}^{-1} = 1 + \frac{{}_1\lambda_{1;\mathbf{t}'0} - {}_1\lambda_{1;\mathbf{t}0}}{{}_1\lambda_{0;\mathbf{t}'0} - {}_1\lambda_{0;\mathbf{t}0}} \tag{6.2'}$$

shows that $0 < {}_1z_{\mathbf{dc}} < 1$ as expected. Since ${}_1\lambda_{0;\mathbf{b}0} < 0$, ${}_1\lambda_{1;\mathbf{b}0} > {}_1\lambda_{1;\mathbf{c}0}$ for ${}_1c_0 < 1$ and ${}_1\lambda_{1;\mathbf{a}0} < 0$, ${}_1\lambda_{0;\mathbf{a}0} > {}_1\lambda_{0;\mathbf{c}0}$ for ${}_1c_0 > 1$ the numerator and denominator of the fraction in the right-hand side of (6.2′) have opposite signs for $\mathbf{t} = \mathbf{c}$, $\mathbf{t}' = \breve{\mathbf{t}}_-$ which confirms that root ${}_1z_{\mathbf{c}\breve{\mathbf{t}}_-}$ lies outside the quantization interval [0, 1]. In addition constraint (A.6) in Appendix A selects the range of the RI $\mu_o$ where the regular basic GS solution lies below the irregular solution $\mathbf{d}0$ provided that the *r*-GRef $V[z \mid {}_1\breve{\mathcal{G}}^{210}_{\downarrow\mathbf{c}0}(z_T)]$ has the discrete energy spectrum. Under the cited restrictions the signed ExpDiffs of the seed solutions $\breve{\mathbf{t}}_-0$ and $\mathbf{d}0$ thus satisfy the inequalities $|{}_1\lambda_{r;\breve{\mathbf{t}}_-0}| > -{}_1\lambda_{r;\mathbf{d}0}$ for $r = 0$ and 1. Since ${}_1\lambda_{0;\breve{\mathbf{t}}_-0}$ and ${}_1\lambda_{1;\breve{\mathbf{t}}_-0}$ have opposite signs we conclude that the root ${}_1z_{\mathbf{d}\breve{\mathbf{t}}_-}$ must lie outside the quantization interval [0, 1] as far as the RIs $\lambda_o$ and $\mu_o$ are restricted as stated above.

The first-order polynomial $z - z_{\mathbf{tt}'}$ satisfies Heun equation (5.5) at the energy $\varepsilon = {}_1\varepsilon_{\mathbf{t}'0}$ iff

$$2B_2[z; {}_1\varepsilon_{\mathbf{t}'0} \mid {}_1\mathcal{G}^{210}_{\downarrow\mathbf{t}0}(z_T); \bar{\sigma}_{\mathbf{t}'}] + C_1[z; {}_1\varepsilon_{\mathbf{t}'0} \mid {}_1^1\mathcal{G}^{210}_{\mathbf{t}0}(z_T); \bar{\sigma}_{\mathbf{t}'}](z - z_{\mathbf{tt}'}) = 0 \tag{6.3}$$

which implies that second-order polynomial (5.4) must be divisible by $z - z_{\mathbf{tt}'}$ at the energy

$\varepsilon = {}_1\varepsilon_{\dagger'0}$ . It is essential that the second-order polynomial in question is not affected by the choice of the basic GS solution $\dagger 0$ and therefore it must be divisible by both first-order polynomials $z - z_{\dagger\dagger'}$ and $z - z_{\dagger\dagger''}$ ($\dagger'' \neq \dagger, \dagger'$):

$$B_2[z; {}_1\varepsilon_{\dagger'0} \mid {}_1\mathcal{G}^{210}_{\downarrow\dagger 0;\dagger''0}(z_T); \bar{\sigma}_{\dagger'}] = \tfrac{1}{2}(\mu_{\dagger'0} + 1)(z - z_{\dagger\dagger'})(z - z_{\dagger''\dagger'}) . \qquad (6.4)$$

The explicit proof of this relation is a little cumbersome so that we moved it to Appendix A.

One can use the CLDT with FF (6.1*) to construct a new rational Liouville potential quantized via the GS Heun polynomials. Namely, making use of (3.21), (6.7), and (6.8) in [2], with $\iota = 1$, $\dagger_1 = \dagger$, $\dagger_2 m_2 = \dagger' 0$, and ${}_\iota\bar{\xi}_{\dagger_1 0;\dagger_2 m_2} = z_{\dagger\dagger'}$, we can represent the RefPF for the PFB ${}^2_1\mathcal{G}^{210}_{\dagger 0;\dagger'0}(z_T)$ as

$$I^o[z \mid {}^2_1\mathcal{G}^{210}_{\dagger 0;\dagger'0}(z_T)] = \frac{1-\lambda_o^2}{4z^2} + \frac{1}{4(z-1)^2} - \frac{2}{(1-z_{\dagger\dagger'})^2} + \frac{O_1[z \mid {}^2_1\mathcal{G}^{210}_{\dagger 0;\dagger'0}(z_T)]}{4z(z-1)(z-z_{\dagger\dagger'})}, \qquad (6.5)$$

where we also took into account that the PF $Q[z; z_{\dagger\dagger'}]$ does not contain the first-order pole if $n_{\dagger m} = 1$:

$$Q[z; z_{\dagger\dagger'}] = -\frac{1}{(1-z_{\dagger\dagger'})^2} . \qquad (6.6)$$

The numerator of the PF with the first-order poles in the right-hand side of (6.5) is defined via (6.30b′) in [2]:

$$O_1[z \mid {}^2_1\mathcal{G}^{210}_{\dagger 0;\dagger'0}(z_T)] = -\left[{}_1d\,{}_1\varepsilon_{\dagger'0} + 4\,{}_1C^0_0(-{}_1\lambda_{0;\dagger'0}, -{}_1\lambda_{1;\dagger'0})\right](z - z_{\dagger\dagger'}) + 8\,\hat{B}_1[z; 0, 1; -{}_1\bar{\lambda}_{\dagger'0}] \qquad (6.7).$$

Removing the energy ${}_1\varepsilon_{\dagger'0}$ from the right-hand side via the relation

$$C_0 \mid {}_1\mathcal{G}^{K\Im 0}_{\dagger'0}; \bar{\sigma}_{\dagger'}] = \tfrac{1}{4}({}_1O^o_0 + {}_1d\,{}_1\varepsilon_{\dagger'0}) + {}_1C^0_0({}_1\lambda_{0;\dagger'0}, {}_1\lambda_{1;\dagger'0}) = 0 \qquad (6.8)$$

and taking advantage of the identity

$${}_1C^0_0(\lambda_0, \lambda_1) - {}_1C^0_0(-\lambda_0, -\lambda_1) = \lambda_0 + \lambda_1 \qquad (6.9)$$

we can express the right-hand side of (6.5) in terms of RefPF for the *r*-GRefPF beam:

$$I^{o}[z\,|\,{}^{2}_{1}G^{210}_{\dagger 0;\dagger' 0}(z_T)] = I^{o}[z\,|\,{}_{1}G^{210}_{\downarrow\dagger 0;\dagger' 0}(z_T)] - \frac{2}{(1-{}_{1}z_{\dagger\dagger'})^2} + \frac{\Delta O_1[z\,|\,{}^{2}_{1}G^{210}_{\dagger 0;\dagger' 0}(z_T)]}{4z(z-1)(z-{}_{1}z_{\dagger\dagger'})}, \tag{6.10}$$

where

$$\Delta\hat{O}^{\downarrow}_{1}[z\,|\,{}^{2}_{1}G^{210}_{\dagger 0;\dagger' 0}(z_T)] = 4[2z-(\mu_{\dagger' 0}-1)z_{\dagger\dagger'} + {}_{1}\lambda_{0;\dagger' 0}-1] \tag{6.11$'$}$$

or alternatively

$$\Delta\hat{O}^{\downarrow}_{1}[z\,|\,{}^{2}_{1}G^{2\Im 0}_{\dagger' 0;\dagger 0}] = 4[2z-(\mu_{\dagger 0}-1)z_{\dagger\dagger'} + {}_{1}\lambda_{0;\dagger 0}-1]. \tag{6.11}$$

Making use of (6.2) one can directly verify that both formulas (6.11′) and (6.11) are equivalent and also represent this polynomial in the following symmetric form:

$$\Delta\hat{O}^{\downarrow}_{1}[z\,|\,{}^{2}_{1}G^{2\Im 0}_{\dagger 0;\dagger' 0}] = 2[4z-(\mu_{\dagger 0}+\mu_{\dagger' 0}-2)z_{\dagger\dagger'} + {}_{1}\lambda_{0;\dagger 0} + {}_{1}\lambda_{1;\dagger' 0}-2] \tag{6.11*}$$

which is nothing but one half of sum of the latter formulas. Note that the derived expressions are applicable to both cases $\Im = 1$ and 2. [The denominator of the fraction in the right-hand side of (6.2) can vanish for $K=\Im=1$ so that this limiting case requires a separate consideration.]

The appropriate RLP generated using the DRtTp ($\Im = 1$) can be thus represented as

$$V[z\,|\,{}^{2}_{1}G^{210}_{\dagger 0;\dagger' 0}(z_T)] = V[z\,|\,{}_{1}G^{210}_{\downarrow\dagger 0;\dagger' 0}(z_T)] + \frac{8z^2(z-1)^2}{(z-z_T)^2(z-z_{\dagger\dagger'})^2} - \frac{z(z-1)\Delta\hat{O}^{\downarrow}_{1}[z\,|\,{}^{2}_{1}G^{210}_{\dagger 0;\dagger' 0}(z_T)]}{(z-z_T)^2(z-z_{\dagger\dagger'})}. \tag{6.12}$$

Again, if $z_T = 2$ the potential is explicitly expressible in terms of x:

$$V[z(x)\,|\,{}^{2}_{1}G^{210}_{\dagger 0;\dagger' 0}(2)] = V[z(x)\,|\,{}_{1}G^{210}_{\downarrow\dagger 0;\dagger' 0}(2)] + \frac{8z^2(x)[z(x)-1]^2}{[z(x)-2]^2[z(x)-z_{\dagger\dagger'})]^2} - \frac{z(x)[z(x)-1]\,\Delta O^{\downarrow}_{1}[z(x)\,|\,{}^{2}_{1}G^{210}_{\dagger 0;\dagger' 0}(2)]}{[z(x)-2]^2[z(x)-z_{\dagger\dagger'})]}, \tag{6.12*}$$

where the change of variable z(x) is an elementary function defined via (4.2).

The energy-dependent gauge transformation,

$$\Phi[z;\varepsilon\,|\,{}^{2}_{1}\mathcal{G}^{210}_{\dagger 0;\dagger' 0}] = z^{\rho_0(\varepsilon|_1\mathcal{G}_{\downarrow\dagger 0;\dagger' 0};\bar{\sigma})}(1-z)^{\rho_1(\varepsilon|_1\mathcal{G}_{\downarrow\dagger 0;;\dagger' 0};\bar{\sigma})}(z-z_{\dagger\dagger'})^{-1} \times \mathrm{Hf}[z;\varepsilon\,|\,{}^{2}_{1}\mathcal{G}^{210}_{\dagger 0;\dagger' 0};\bar{\sigma}], \tag{6.13}$$

converts the RCSLE with the RefPF (6.10) into the Heun equation

$$\Big\{z(z-1)(z-z_{\dagger\dagger'})\frac{d^2}{dz^2} + 2B_2[z;\bar{\rho}(\varepsilon\,|\,{}_1\mathcal{G}^{210}_{\downarrow\dagger 0;\dagger' 0};\bar{\sigma});z_{\dagger\dagger'}]\frac{d}{dz} + C_1[z;\varepsilon\,|\,{}^{2}_{1}\mathcal{G}^{210}_{\dagger 0;\dagger' 0};\bar{\sigma}]\Big\}\mathrm{Hf}[z;\varepsilon\,|\,{}^{2}_{1}\mathcal{G}^{210}_{\dagger 0;\dagger' 0};\bar{\sigma}] = 0. \tag{6.14}$$

Similarly to (5.5) the second-order polynomial in the left-hand side of this differential equation is defined via (5.6), with $z_T$ changed for $z_{\dagger\dagger'}$. As for the free term it is given by (5.7) with ${}^{1}_{1}\mathcal{G}^{210}_{\dagger 0}$ substituted for ${}^{2}_{1}\mathcal{G}^{210}_{\dagger 0;\dagger' 0}$. Likewise the energy-dependent accessory parameter is can be obtained from (5.10) by formally changing $z_T$ and ${}^{1}_{1}\mathcal{G}^{210}_{\dagger 0}$ for $z_{\dagger\dagger'}$ and ${}^{2}_{1}\mathcal{G}^{210}_{\dagger 0;\dagger' 0}$ respectively:

$$q(\varepsilon\,|\,{}^{2}_{1}\mathcal{G}^{210}_{\dagger 0;\dagger' 0};\bar{\sigma}) \equiv -C_{1;0}\,|\,\varepsilon\,|\,{}^{2}_{1}\mathcal{G}^{210}_{\dagger 0;\dagger' 0};\bar{\sigma}) = -\tfrac{1}{4}O_{1;0}\,|\,{}^{2}_{1}\mathcal{G}^{210}_{\dagger 0;\dagger' 0}) + \tfrac{1}{4}\,{}_1d\,\varepsilon - 2\rho_0(\varepsilon\,|\,{}_1\mathcal{G}^{210}_{\downarrow\dagger 0;\dagger' 0};\bar{\sigma})\rho_1(\varepsilon\,|\,{}_1\mathcal{G}^{210}_{\downarrow\dagger 0;\dagger' 0};\bar{\sigma})z_{\dagger\dagger'} - 2\rho_0(\varepsilon\,|\,{}_1\mathcal{G}^{210}_{\downarrow\dagger 0;\dagger' 0};\bar{\sigma}). \tag{6.15}$$

Though two Heun problems look much the same a crucial difference comes from the fact that, the position of outer singular point (6.2) depends on values of the RIs $\lambda_o$ and $\mu_o$. For this reason we refer to RLP (6.12) as conditionally exactly solvable by Heun polynomials'(Hp-CEQ). It was Takemura [29] who first brought attention to the Heun equation of this nontrivial type in connection with Quesne's analysis [30, 31] of the SUSY partner of the isotonic oscillator quantized by orthogonal $X_1$–Jacobi polynomials [32, 33]. As emphasized by the author [16] one can construct similar Hp-CEQ potentials by applying the CLDT with the nodeless FF **a**1 or **b**1 (if exists) to any shape-invariant *r*-GRef potential.

By representing (8.10a) and (8.10b) in [2] as

$$\alpha(\varepsilon\,|\,{}^{1}_{1}G^{210}_{\mathbf{t}0\downarrow\mathbf{t}'0};\bar{\sigma}) = \alpha(\varepsilon\,|\,{}^{2}_{1}G^{210}_{\mathbf{t}0;\mathbf{t}'0};\bar{\sigma}) + n - n_{*\mathbf{t}'0} - 1 \qquad (6.16a)$$

and

$$\beta(\varepsilon\,|\,{}^{1}_{1}G^{210}_{\mathbf{t}0\downarrow\mathbf{t}'0};\bar{\sigma}) = \beta(\varepsilon\,|\,{}^{2}_{1}G^{210}_{\mathbf{t}0;\mathbf{t}'0};\bar{\sigma}) + n - n_{*\mathbf{t}'0} - 1, \qquad (6.16b)$$

where

$$*\mathbf{t}' = \mathbf{b}, \mathbf{a}, \mathbf{d}, \mathbf{c} \text{ for } \mathbf{t}' = \mathbf{a}, \mathbf{b}, \mathbf{c}, \mathbf{d}, \qquad (6.17)$$

and making use of Suzko's reciprocal formula [23-25]

$$*\phi[z\,|\,{}^{1}_{1}G^{210}_{\mathbf{t}0\downarrow\mathbf{t}'0};\mathbf{t}'0] = \frac{2z(1-z)}{(z-z_T)\,\phi[z\,|\,{}^{1}_{1}G^{210}_{\mathbf{t}0\downarrow\mathbf{t}'0};\mathbf{t}'0]} \qquad (6.18)$$

$$\propto \frac{\sqrt{z(1-z)}}{z - {}_1z_{\mathbf{t}\mathbf{t}'}}\, z^{-\frac{1}{2}\,{}_1\lambda_{0;\mathbf{t}'0}}\,(1-z)^{-\frac{1}{2}\,{}_1\lambda_{1;\mathbf{t}'0}} \qquad (6.18^*)$$

for the FF of the inverse CLDT from the PFB ${}^{2}_{1}G^{210}_{\mathbf{t}0;\mathbf{t}'0}$ to the PFB ${}^{1}_{1}G^{210}_{\mathbf{t}0\downarrow\mathbf{t}'0}$ we find that n and $n_{*\mathbf{t}0}$ in (6.16a) and (6.16b) are equal to 1 and 0, respectively, so that

$$\alpha(\varepsilon\,|\,{}^{1}_{1}G^{210}_{\mathbf{t}0;\mathbf{t}'0};\bar{\sigma}) = \alpha(\varepsilon\,|\,{}^{2}_{1}G^{210}_{\mathbf{t}0;\mathbf{t}'0};\bar{\sigma}) \qquad (6.19a)$$

and

$$\beta(\varepsilon\,|\,{}^{1}_{1}G^{210}_{\mathbf{t}0;\mathbf{t}'0};\bar{\sigma}) = \beta(\varepsilon\,|\,{}^{2}_{1}G^{210}_{\mathbf{t}0;\mathbf{t}'0};\bar{\sigma})\,. \qquad (6.19b)$$

As a direct consequence of this result we conclude that the CLDT with FF (6.1) acting upon the Heun polynomial $Hp_{m''+1}[z|{}^{1}_{1}G^{210}_{\mathbf{t}0\downarrow\mathbf{t}'0;\mathbf{t}''m''};\mathbf{t}''m'']$ keeps its order unchanged.

## 7. Conclusions and further developments

The main purpose of this is to illustrate main features of the general theory of SUSY ladders of RCSLEs with second-order poles [2] using the DRtTP *r*-GRef potential as a starting point. A distinctive attribute of this potential is that energies of seed solutions used to construct its rational SUSY partners (including the eigenvalues) can be obtained by solving a cubic equation. It thus presents an intermediate case between shape-invariant Rosen-Morse and generic *r*-GRef potentials with factorization energies given by quadratic and quartic equations respectively.

The appropriate Liouville transformation transforms the RCSLE with the *r*-GRef Bose invariant into the Schrödinger equation in the dimension x, with the potential generally defined as an implicit function of x [1]. The DKV potential [11] generated by means of the TP with the double root of 2 represents a remarkable exception. Namely, both potential and all the eigenfunctions the given Schrödinger equation can be also expressed in terms of elementary transformation functions, as a counter-example to the recent statement in [12] that this is the prerogative of shape-invariant potentials.

Another distinctive attribute of the DRtTP *r*-GRef potential is that its SUSY partners generated using basic FFs is exactly quantized by Heun polynomials. In particular this is true for the DRtTP reduction of the CGK potential [34] which is constructed using the ground-energy eigenfunction as the FF for the single-step DT.

In Section 6 we also proved that double-step SUSY partners of the DRtTP *r*-GRef potential is generated using basic seed solutions is conditionally exactly quantized by Heun polynomials. The distinction in the terminology comes from the fact that the position of the outer singular point depends on values of the RIs. However, as pointed to in [16] and illuminated in detail in a separate paper [35] this is the common feature of all the *r*-GRef potentials on the line.

In [35] we also analyze multi-step DTs of the generic *r*-GRef potential regular Gauss seed solutions **†**m below ground-energy level coupled with pairs **c**,v;**c**,v+1 of juxtaposed eigenfunctions. It was Bagrov and Samsonov [36, 37] who first used such pairs to construct nodeless two-step FFs based on the Krein theorem [38]. In particular they discuss two-step SUSY partner of the Morse potential [39] implicitly assuming that the theorem (formulated by Krein for potentials quantized on either finite interval or half-line) can be automatically extended to potentials on the line. It is worth stressing in this connection that Adler's paper [40] published in the same time was solely restricted to a finite quantization interval. Nevertheless it is usually taken for granted [36, 37, 12, 26] that the Krein-Adler theorem also covers potentials on the line. In [35] we extend Adler's arguments [40] to the Sturm-Liouville problems quantized on a finite interval. Our proof automatically covers SUSY partners of the generic *r*-GRef potential quantized on the finite interval $0 \le z \le 1$ which includes their DRtTP reductions of our current interest.

As two roots of the TP merge the generic *r*-GRef potential turns into DRtTP potentials conditionally exactly quantized by GS Heine polynomials. Again, choosing the TP double root equal to 2 makes it possible to represent each potential and its eigenfunctions as transcendental elementary functions of x.

## Appendix A

**First-order polynomial solutions of the Heun equations associated with basic DRtTP sibling beams**

Let us explicitly prove that the polynomial

$$B_2[z; 0, 1, z_T; {}_1\rho_{0;\dagger'0}, {}_1\rho_{1;\dagger'0}, -1]$$

$$= [{}_1\rho_{0;\dagger'0}(z-1) + {}_1\rho_{1;\dagger'0} z](z - z_T) - z(z-1) \tag{A.1}$$

$$= \tfrac{1}{2}[(\mu_{\dagger'0} - 1)z^2 - [z_T(\mu_{\dagger'0} + 1) + {}_1\lambda_{0;\dagger'0} - 1]z + z_T({}_1\lambda_{0;\dagger'0} + 1) \tag{A.1'}$$

can be decomposed as

$$B_2[z; 0, 1, z_T; {}_1\rho_{0;\dagger'0} + 1, {}_1\rho_{1;\dagger'0} + 1, -1] = -\frac{1}{2}(z - z_{\dagger\dagger'})C_1[z; {}_1\varepsilon_{\dagger'0} \mid {}^{1}_{1}\boldsymbol{G}^{210}_{\dagger 0\downarrow\dagger'0}; \bar{\sigma}_{\dagger'}]\,. \tag{A.2}$$

As mentioned in Section 6 this decomposition is the necessary and sufficient condition for the polynomial $z - z_{\dagger\dagger'}$ to satisfy Heun equation (5.5). The first-order polynomial

$$C_1[z; {}_1\varepsilon_{\dagger'0} \mid {}^1_1\mathcal{G}^{210}_{\dagger0\downarrow\dagger'0}(z_T); \bar{\sigma}_{\dagger'}] = \tfrac{1}{4}\Delta\hat{O}^{\downarrow}_1[z \mid {}^1_1\mathcal{G}^{210}_{\dagger0\downarrow\dagger'0}(z_T)] - 2\hat{B}_1[\xi; 0, 1; \bar{\lambda}_{\dagger'0}]\} \qquad (A.3)$$

in the right-hand side of (A.2) is obtained from (3.29) in [2] by setting ${}^{\ell}_{\iota}\mathbf{B}^{K\Im 0} = {}^1_1\mathcal{G}^{210}_{\dagger0}(z_T)$. Since n=0 most terms in the right-hand side of the latter formula disappear so that

$$C_1[z; {}_1\varepsilon_{\dagger'0} \mid {}^1_1\mathcal{G}^{210}_{\dagger0\downarrow\dagger'0}; \bar{\sigma}_{\dagger'}] = [\tfrac{1}{4}\,{}_1d({}_1\varepsilon_{\dagger'0} - {}_1\varepsilon_{\dagger0}) + 2\rho_{0;\dagger'0}\,\rho_{1;\dagger'0}](z - z_T)$$
$$-(\mu_{\dagger0} + \mu_{\dagger'0})z + {}_1\lambda_{0;\dagger0} + {}_1\lambda_{0;\dagger'0}. \qquad (A.3^*)$$

Substituting

$${}_1d\,{}_1\varepsilon_{0;\dagger'0} + 8\,{}_1\rho_{0;\dagger'0}\,{}_1\rho_{1;\dagger'0} + {}_1O^0_0 = 0 \qquad (A.4)$$

and

$$\hat{O}^{\downarrow}_1[z \mid {}^1_1\mathcal{G}^{210}_{\dagger0}] = {}_1O^0_0(z - z_T) + \Delta\hat{O}^{\downarrow}_1[z \mid {}^1_1\mathcal{G}^{210}_{\dagger0} \qquad (A.5)$$

into (A.3*) then directly leads to (A.2), where the first-order polynomials in the right-hand side are defined via (5.2′) in Section 5 and (6.17) in [2], with ι set to 1, i.e.,

$$\hat{B}_1[z; 0, 1; \bar{\lambda}_{\dagger'0}] = \tfrac{1}{2}[(\mu_{\dagger'0} + 1)z - {}_1\lambda_{0;\dagger'0} - 1]. \qquad (A.6)$$

Substituting the cited expressions into (A.3) we can thus represent the latter as

$$C_1[z; {}_1\varepsilon_{\dagger'0} \mid {}^1_1\mathcal{G}^{210}_{\dagger0\downarrow\dagger'0}(z_T); \bar{\sigma}_{\dagger'}] = (\mu_{\dagger0} - 1)(z - z_T) - (\mu_{\dagger0} + \mu_{\dagger'0} - 2)z + {}_1\lambda_{0;\dagger0} + {}_1\lambda_{0;\dagger'0}. \qquad (A.7)$$

One can directly verify that the leading coefficients of the polynomials representing the left- and right-sides of (A.1) do coincide as expected.

Our next step is to prove that polynomial (A.2) vanishes at $z = z_{\dagger\dagger'}$. Combining (7.15) with (2.17) and (2.18) written for both pairs of the signed ExpDiffs ${}_1\lambda_{r;\dagger0}$ and ${}_1\lambda_{r;\dagger'0}$ ($r$ = 0, 1) one finds

$$z_T^2(\mu^2_{\dagger'0} - \mu^2_{\dagger0}) = {}_1\lambda^2_{0;\dagger'0} - {}_1\lambda^2_{0;\dagger0} \qquad (A.8a)$$

and

$$(1-z_T)^2(\mu^2_{\dagger'0}-\mu^2_{\dagger 0}) = {}_1\lambda^2_{1;\dagger'0} - {}_1\lambda^2_{1;\dagger 0}. \qquad \text{(A.8b)}$$

Making use of (6.2a) and (6.2b) we can transform these quadratic formulas into the linear constraints on $\mu_{\dagger'0}$, $\mu_{\dagger 0}-\mu_{\dagger'0}$, and the appropriate ChExp ${}_1\rho_{0;\dagger'0}$ or ${}_1\rho_{1;\dagger'0}$:

$$2z_T^2\mu_{\dagger'0} - 2z_{\dagger\dagger'}(2\,{}_1\rho_{0;\dagger'0}-1) = (z_{\dagger\dagger'}-z_T)(z_{\dagger\dagger'}+z_T)(\mu_{\dagger 0}-\mu_{\dagger'0}) \qquad \text{(A.9a)}$$

and

$$(1-z_T)^2\mu_{\dagger'0} - (1-z_{\dagger\dagger'})(2\,{}_1\rho_{1;\dagger'0}-1) = (z_{\dagger\dagger'}-z_T)(z_{\dagger\dagger'}+z_T-2)(\mu_{\dagger 0}-\mu_{\dagger'0}). \qquad \text{(A.9b)}$$

Multiplying the latter constraints respectively by $z_{\dagger\dagger'}+z_T-2$ and $z_{\dagger\dagger'}+z_T$ and subtracting the resultant expressions we then exclude $\mu_{\dagger 0}-\mu_{\dagger'0}$. This brings us to the sought-for constraint

$$(z_{\dagger\dagger'}-z_T)[(z_{\dagger\dagger'}-1)\,{}_1\rho_{0;\dagger'0} + z_{\dagger\dagger'}\,{}_1\rho_{1;\dagger'0}] - z_{\dagger\dagger'}(z_{\dagger\dagger'}-1) = 0 \qquad \text{(A.10)}$$

which is nothing but the requirement for polynomial (A.1) to have zero at z = $z_{\dagger\dagger'}$:

$$B_2[z_{\dagger\dagger'};0,1,z_T;\,{}_1\rho_{0;\dagger'0},\,{}_1\rho_{1;\dagger'0},-1] = 0. \qquad \text{(A.10*)}$$

It directly follows from this condition that polynomial (A.1) can be decomposed as

$$B_2[z;0,1,z_T;\,{}_1\rho_{0;\dagger 0},\,{}_1\rho_{1;\dagger 0},-1] = \tfrac{1}{2}(\mu_{\dagger 0}-1)(z-z_{\dagger\dagger'})(z-z_{\dagger\dagger''}), \qquad \text{(A.11)}$$

where we changed ${}_1\rho_{r;\dagger'0}$ for ${}_1\rho_{r;\dagger 0}$ to arrange subscripts in a more symmetric fashion. Substituting (A.1) into the left-hand side of (A.2) and subtracting (A.1′) we come to (A.7) which completes the proof.

**Appendix B**

**Regular AEH solutions below the ground energy levels of Hp-EQ SUSY partners of the DRtTP *r*-GRef potential**

The regular AEH solution $\phi_{\breve{t}_m}[z \mid {}^{1}_{1}\breve{G}^{210}_{\breve{t}0\downarrow cm}(z_T)]$ of the RCSLE with the Bose invariant $I[z;\varepsilon \mid {}^{1}_{1}\breve{G}^{210}_{\breve{t}0\downarrow cm}(z_T)]$ lies below the ground energy, ${}_1\varepsilon_{c0}$, ${}_1\varepsilon_{c1}$, or ${}_1\varepsilon_{d0}$, of the basic sibling potential $V[z \mid {}^{1}_{1}\breve{G}^{210}_{\breve{t}0\downarrow cm}(z_T)]$ iff

$$4\sqrt{{}_1c_0}\,|g_m(\mu_o)| > |1-\sqrt{{}_1c_0}\,|[\sqrt{{}_1\breve{\Delta}_0(\mu_o;{}_1c_0)} - 2(\sqrt{{}_1c_0}+1)](2m+1) \text{ for } m>0 \quad \text{(B.1)}$$

$$4\sqrt{{}_1c_0}\,|g_m(\mu_o)| > |1-\sqrt{{}_1c_0}\,|[\sqrt{{}_1\breve{\Delta}_1(\mu_o;{}_1c_0)} - 6(\sqrt{{}_1c_0}+1)](2m+1), \quad \text{(B.1c)}$$

and

$$4\sqrt{{}_1c_0}\,|g_m(\mu_o)| > |1-\sqrt{{}_1c_0}\,|[\sqrt{{}_1\breve{\Delta}_0(\mu_o;{}_1c_0)} + 2(\sqrt{{}_1c_0}+1)](2m+1) \quad \text{(B.1d)}$$

for $\breve{t} = \breve{t}_$, $c$, and $d$ respectively. Condition (B.1) is equivalent to the requirement that the seed solution $\breve{t}_m$ is nodeless. One can also use any nodeless seed solution $\breve{t}_m$ to construct the double-step Hi-CEQ SUSY partner $V[z \mid {}^{2}_{1}\breve{G}^{210}_{c0;\breve{t}_m\downarrow cm}(z_T)]$ of the DRtTP *r*-GRef potential $V[z \mid {}_1\breve{G}^{210}_{\downarrow cm}(z_T)]$. In addition any regular seed solution with zeros between 0 and 1 is equally acceptable as far as it satisfies (B.1.c). On other hand, to construct the double-step Hi-CEQ SUSY partner $V[z \mid {}^{2}_{1}\breve{G}^{210}_{d0;\breve{t}_m\downarrow cm}(z_T)]$ of the DRtTP *r*-GRef potential one can use only a subset of nodeless seed solutions selected via (B.1d).

By expressing $\mu_o$ in the right-hand side of the listed inequalities in terms of discriminant (2.27) with m set to either 0 or 1:

$$4\sqrt{{}_1c_0}\,\mu_o^2 = \tfrac{1}{4}\,{}_1\breve{\Delta}_0(\mu_o;{}_1c_0) - (\sqrt{{}_1c_0}-1)^2 \quad \text{(B.2)}$$

or

$$4\sqrt{{}_1c_0}\,\mu_o^2 = \tfrac{1}{4}\,{}_1\breve{\Delta}_1(\mu_o;{}_1c_0) - 9(\sqrt{{}_1c_0} - 1)^2 \quad \text{(B.2c)}$$

one can represent the given conditions as

$$M_2[m;\mu_o;{}_1c_0\,|\breve{\mathbf{t}}_-] \equiv 4m^2 + 2[\,|1-\sqrt{{}_1c_0}\,|\sqrt{{}_1\breve{\Delta}_0(\mu_o;{}_1c_0)} + 2 - 2|1-{}_1c_0\,|\,]m + M_2[0;\mu_o;{}_1c_0\,|\breve{\mathbf{t}}_-] < 0 \quad \text{(B.3)}$$

$$M_2[m;\mu_o;{}_1c_0\,|\mathbf{c}] \equiv 4m^2 + 2[\,|1-\sqrt{{}_1c_0}\,|\sqrt{{}_1\breve{\Delta}_1(\mu_o;{}_1c_0)} + 2 - 6|1-{}_1c_0\,|\,]m + M_2[0;\mu_o;{}_1c_0\,|\mathbf{c}] < 0 \quad \text{(B.3c)}$$

and

$$M_2[m;\mu_o;{}_1c_0\,|\mathbf{d}] \equiv 4m^2 + 2[\,|1-\sqrt{{}_1c_0}\,|\sqrt{{}_1\breve{\Delta}_1(\mu_o;{}_1c_0)} + 2|1-{}_1c_0\,| + 2\,]m + M_2[0;\mu_o;{}_1c_0\,|\mathbf{c}] < 0 \quad \text{(B.3d)}$$

if $2m+1 < \mu_o$ or

$$\tilde{M}_2[m;\mu_o;{}_1c_0\,|\breve{\mathbf{t}}_-] \equiv 4m^2 - 2[\,|1-\sqrt{{}_1c_0}\,|\sqrt{{}_1\breve{\Delta}_0(\mu_o;{}_1c_0)} + 2 + 2|1-{}_1c_0\,|\,]m + \tilde{M}_2[0;\mu_o;{}_1c_0\,|\breve{\mathbf{t}}_-] > 0 \quad \text{(B.3′)}$$

$$\tilde{M}_2[m;\mu_o;{}_1c_0\,|\mathbf{c}] \equiv 4m^2 - 2[\,|1-\sqrt{{}_1c_0}\,|\sqrt{{}_1\breve{\Delta}_1(\mu_o;{}_1c_0)} - 2 - 6|1-{}_1c_0\,|\,]m + \tilde{M}_2[0;\mu_o;{}_1c_0\,|\mathbf{c}] > 0 \quad \text{(B.3c′)}$$

and

$$\tilde{M}_2[m;\mu_o;{}_1c_0\,|\mathbf{d}] \equiv 4m^2 + 2[\,|1-\sqrt{{}_1c_0}\,|\sqrt{{}_1\breve{\Delta}_1(\mu_o;{}_1c_0)} + 2|1-{}_1c_0\,| + 2\,]m + \tilde{M}_2[0;\mu_o;{}_1c_0\,|\mathbf{c}] > 0 \quad \text{(B.3d′)}$$

if $2m+1 > \mu_o$. As for the free terms of the quadratic polynomials $M_2[m;\mu_o;{}_1c_0\,|\mathbf{t}]$ and $\tilde{M}_2[m;\mu_o;{}_1c_0\,|\mathbf{t}]$ in m we in turn represent them as quadratic polynomials in $\tfrac{1}{2}\sqrt{{}_1\breve{\Delta}_\kappa(\mu_o;{}_1c_0)}$, , namely, we set

$$M_2[0;\mu_o;{}_1c_0\,|\breve{\mathbf{t}}_-] \equiv -N_2[\tfrac{1}{2}\sqrt{{}_1\breve{\Delta}_0(\mu_o;{}_1c_0)};{}_1c_0\,|\breve{\mathbf{t}}_-], \quad \text{(B.4)}$$

$$M_2[0;\mu_o;{}_1c_0\,|\mathbf{c}] \equiv -N_2[\tfrac{1}{2}\sqrt{{}_1\breve{\Delta}_1(\mu_o;{}_1c_0)};{}_1c_0\,|\mathbf{c}], \quad \text{(B.4c)}$$

$$M_2[0;\mu_o;{}_1c_0\,|\mathbf{d}] \equiv -N_2[{}^1\!/_2\sqrt{{}_1\breve{\Delta}_0(\mu_o;{}_1c_0)};{}_1c_0\,|\mathbf{d}], \tag{B.4d}$$

and

$$\tilde{M}_2[0;\mu_o;{}_1c_0\,|\breve{\mathbf{t}}_-] \equiv -\tilde{N}_2[{}^1\!/_2\sqrt{{}_1\breve{\Delta}_0(\mu_o;{}_1c_0)};{}_1c_0\,|\breve{\mathbf{t}}_-], \tag{B.4′}$$

$$\tilde{M}_2[0;\mu_o;{}_1c_0\,|\mathbf{c}] \equiv -\tilde{N}_2[{}^1\!/_2\sqrt{{}_1\breve{\Delta}_1(\mu_o;{}_1c_0)};{}_1c_0\,|\mathbf{c}], \tag{B.4c′}$$

$$\tilde{M}_2[0;\mu_o;{}_1c_0\,|\mathbf{d}] \equiv -\tilde{N}_2[{}^1\!/_2\sqrt{{}_1\breve{\Delta}_0(\mu_o;{}_1c_0)};{}_1c_0\,|\mathbf{d}], \tag{B.4d′}$$

where

$$N_2[\delta;{}_1c_0\,|\breve{\mathbf{t}}_-] \equiv \delta^2 - 2\,|1-\sqrt{{}_1c_0}\,|\delta + 2\,|\,{}_1c_0 - 1| - (\sqrt{{}_1c_0} - 1)^2 - 1, \tag{B.5}$$

$$N_2[\delta;{}_1c_0\,|\mathbf{c}] \equiv \delta^2 - 2\,|1-\sqrt{{}_1c_0}\,|\delta + 6\,|\,{}_1c_0 - 1| - 9(\sqrt{{}_1c_0} - 1)^2 - 1, \tag{B.5c}$$

$$N_2[\delta;{}_1c_0\,|\mathbf{d}] \equiv \delta^2 - 2\,|1-\sqrt{{}_1c_0}\,|\sqrt{{}_1\breve{\Delta}_0(\mu_o;{}_1c_0)}\delta - 2\,|\,{}_1c_0 - 1| - (\sqrt{{}_1c_0} - 1)^2 - 1, \tag{B.5d}$$

and

$$N_2[\delta;{}_1c_0\,|\breve{\mathbf{t}}_-] \equiv \delta^2 + 2\,|1-\sqrt{{}_1c_0}\,|\delta - 2\,|\,{}_1c_0 - 1| - (\sqrt{{}_1c_0} - 1)^2 - 1, \tag{B.5′}$$

$$N_2[\delta;{}_1c_0\,|\mathbf{c}] \equiv \delta^2 + 2\,|1-\sqrt{{}_1c_0}\,|\delta - 6\,|\,{}_1c_0 - 1| - 9(\sqrt{{}_1c_0} - 1)^2 - 1, \tag{B.5c′}$$

$$N_2[\delta;{}_1c_0\,|\mathbf{d}] \equiv \delta^2 + 2\,|1-\sqrt{{}_1c_0}\,|\sqrt{{}_1\breve{\Delta}_0(\mu_o;{}_1c_0)}\delta + 2\,|\,{}_1c_0 - 1| - (\sqrt{{}_1c_0} - 1)^2 - 1. \tag{B.5′d}$$

Let us first focus on the quadratic polynomials $M_2[m;\mu_o;{}_1c_0\,|\mathbf{t}]$ in m which are associated with seed solutions $\breve{\mathbf{t}}$_m from the primary sequence ( $\breve{\mathbf{t}}_- = \mathbf{a}$ or $\mathbf{b}$ for ${}_1c_0 > 1$ or ${}_1c_0 < 1$ respectively) and prove that the regular AEH solution $\phi_{\breve{\mathbf{t}}_m}[z\,|\,{}^1_1\breve{G}^{210}_{\breve{\mathbf{t}}0\downarrow\mathbf{c}m}(z_T)]$ does not have nodes inside the quantization interval iff

$$\mu_o > \max\left\{1,\ [\delta^2_{\mathbf{d};+}({}_1c_0) - (\sqrt{{}_1c_0} - 1)^2]/\sqrt{{}_1c_0}\,]\right\}, \tag{B.6}$$

where

$$\delta_{\mathbf{d};+}({}_1c_0) = |1-\sqrt{{}_1c_0}\,| + \sqrt{1 + 2|1 - {}_1c_0\,|(2m+1) + 2(1-\sqrt{{}_1c_0})^2}. \tag{B.7}$$

Indeed condition (B.6) holds for m=0 iff the quadratic polynomial $M_2[m;\mu_o;{}_1c_0\,|\mathbf{d}]$ has a negative free term, i.e., iff

$$N_2[\tfrac{1}{2}\sqrt{\breve{\Delta}_0(\mu_o;{}_1c_0)};{}_1c_0\,|\mathbf{d}] > 0. \tag{B.8}$$

Since the roots $\delta_{\mathbf{d};-}(0;{}_1c_0) < 0$ and $\delta_{\mathbf{d};+}(m;{}_1c_0) > 0$ of polynomial (B.5d) have opposite sign the latter condition holds

$$\breve{\Delta}_0(\mu_o;{}_1c_0) > 4\delta^2_{\mathbf{d};+}({}_1c_0). \tag{B.9}$$

Expressing the latter condition in terms $\mu_o$ via (B.2) we come to (B.6) which completes the proof.

The reader can proceed in a similar way to derive bounds for the discriminants to satisfy inequalities (B.3) and (B.3c). The derivation is straightforward but a little more tedious. As for inequalities (B.3′), (B.3c′), and (B.3d′) associated with regular seed solution $\breve{\mathbf{t}}'_{-}m$ from the secondary sequence ( $\breve{\mathbf{t}}_{-} = \mathbf{b}'$ or $\mathbf{a}'$ for ${}_1c_0 > 1$ or ${}_1c_0 < 1$ respectively). let us only mentioned that the leading coefficients of three quadratic polynomials $\tilde{M}_2[m;\mu_o;{}_1c_0\,|\mathbf{t}]$, where $\mathbf{t} = \breve{\mathbf{t}}_{-}$, $\mathbf{c}$, and $\mathbf{d}$, are all positive so that the appropriate FFs are nodeless at sufficiently large m, in agreement with the asymptotic formulas derived in Section 2.